\RequirePackage{fix-cm}
\documentclass[smallextended]{svjour3}       % onecolumn (second format)
\smartqed  % flush right qed marks, e.g. at end of proof
\usepackage{amssymb}
\usepackage{graphicx}
\usepackage{subfigure}
\usepackage{mathptmx}      % use Times fonts if available on your TeX system
\usepackage{natbib}
\usepackage{aas_macros}
\usepackage{textcomp}
\usepackage{color}
\journalname{Experimental Astronomy}
\begin{document}

\title{XIPE: the X-ray Imaging Polarimetry Explorer%\thanks{Grants or other notes
}

\titlerunning{XIPE}        % if too long for running head

\author{Paolo Soffitta, Xavier Barcons, Ronaldo Bellazzini, Jo\~{a}o Braga, Enrico Costa, George W. Fraser, Szymon Gburek,
 Juhani Huovelin, Giorgio Matt, Mark Pearce, Juri Poutanen, Victor Reglero, Andrea Santangelo, Rashid A. Sunyaev,
 Gianpiero Tagliaferri, Martin Weisskopf, Roberto Aloisio, Elena Amato, Primo Attin\'{a}, Magnus Axelsson, Luca Baldini, Stefano Basso, Stefano Bianchi, Pasquale Blasi, Johan  Bregeon, Alessandro Brez,
 Niccol\'{o} Bucciantini, Luciano Burderi, Vadim Burwitz, Piergiorgio Casella, Eugene Churazov,  Marta Civitani, Stefano Covino, Rui Miguel Curado da Silva,
 Giancarlo Cusumano, Mauro Dadina, Flavio D'Amico, Alessandra De Rosa, Sergio Di Cosimo, Giuseppe Di Persio, Tiziana Di Salvo, Michal Dovciak, Ronald Elsner,
 Chris J. Eyles, Andrew C. Fabian, Sergio Fabiani, Hua Feng, Salvatore Giarrusso, Ren\'{e} W. Goosmann, Paola Grandi, Nicolas Grosso, Gianluca Israel, Miranda Jackson, Philip Kaaret, Vladimir Karas,
 Michael Kuss, Dong Lai,
 Giovanni La Rosa, Josefin Larsson, Stefan Larsson, Luca Latronico, Antonio Maggio,
 Jorge Maia, Fr\'{e}d\'{e}ric  Marin, Marco Maria Massai, Teresa Mineo, Massimo Minuti, Elena Moretti, Fabio Muleri, Stephen L. O'Dell, Giovanni Pareschi,
 Giovanni Peres,  Melissa Pesce, Pierre-Olivier Petrucci, Michele Pinchera, Delphine Porquet, Brian Ramsey, Nanda Rea, Fabio Reale, Juana Maria Rodrigo, Agata R\'{o}\.{z}a\'{n}ska,
 Alda Rubini, Pawel Rudawy, Felix Ryde, Marco Salvati, Valdivino Alexandre de Santiago J\'{u}nior, Sergey Sazonov, Carmelo Sgr\'{o}, Eric Silver,
 Gloria Spandre, Daniele Spiga, Luigi Stella, Toru Tamagawa, Francesco Tamborra, Fabrizio Tavecchio, Teresa Teixeira Dias, Matthew van Adelsberg,
 Kinwah Wu, Silvia Zane}

 \authorrunning{Soffitta et al.} % if too long for running head

%\institute{P. Soffitta \at
  %            IAPS/INAF Via Fosso del Cavaliere 100 \\
  %            Tel.: +39-06-49934006\\
  %            Fax: +39-06-45488188\\
  %            \email{paolo.soffitta@iaps.inaf.it}           %  \\
           %   \and
           %E. Costa \at
           %   IAPS-Rome/INAF
           %   \and
           %R. Bellazzini \at
           %   INFN-Pisa

           %    \and
           %   Joao Br\~{a}ga \at
           %   INPE S Jos\'e dos Campos, Brazil

              %\and
              %Juhani Huovelin \at
              %University of Helsinki, Finland

%}

\institute{E. Costa, A. De Rosa, S. Di Cosimo, G. Di Persio, S.
Fabiani, F. Muleri, A. Rubini, P. Soffitta \at
              IAPS/INAF, Via Fosso del Cavaliere 100, 00133 Rome, Italy\\
              Tel.: +39-06-49934006 (P. Soffitta)\\
              Fax: +39-06-45488188\\
              \email{paolo.soffitta@iaps.inaf.it}
              \and X. Barcons \at
              Instituto de Física de Cantabria (CSIC-UC)\\
              Avenida de los Castros, s/n, E-39005 Santander Cantabria Spain
              \and R. Bellazzini, J. Bregeon, A. Brez,  L. Latronico, M. Kuss , M. Minuti, M. Pesce, \\
              M. Pinchera, C. Sgr\'{o}, G. Spandre \at
                  INFN-Pisa, Largo B. Pontecorvo 3, 56127 Pisa, Italy
                  \and J. Braga,  F. D'Amico, V. Santiago \at
                  INPE Div de Astrofísica, Av dos Astronautas 1758\\
                  Jd. Granja - CEP: 12227-010, S\~ao Jos\'e dos Campos - SP, Brazil
                  \and G. W. Fraser \at
                  University of Leicester, Space Research Centre\\
                  Department of Physics and Astronomy, Leicester LE1 7RH, United Kingdom
                  \and S. Gburek \at
                  Space Research Centre, Polish Academy of Sciences\\
                  Solar Physics Division, 51-622 Wroclaw, ul. Kopernika, Poland
                  \and J. Huovelin \at
                  Department of Physics
                  Erik Palmenin aukio 1, 00014 University of Helsinki, Finland
                  \and S. Bianchi, G. Matt, F. Tamborra \at
                  Dipartimento di Fisica "E. Amaldi" \\
                  Universita' degli Studi Roma Tre, Via della Vasca Navale 84, 00146 Rome, Italy
                  \and M. Axelsson, M. S. Jackson, J. Larsson, S. Larsson, E. Moretti, M. Pearce,, F. Ryde \at
                  KTH, Royal Institute of Technology\\
                  Department of Physics\\
                  \& the Oskar Klein Centre for Cosmoparticle Physics\\
                  AlbaNova University Centre, Stockholm, Sweden
                  \and J. Poutanen \at
                  Astronomy Division \\
                  Department of Physics, PO Box 3000, FI-90014 University of Oulu, Finland
                  \and V. Reglero , J. M. Rodrigo \at
                  Universidad de Valencia\\
                  Astronomia i Astrofísica\\
                  Instituto de Ciencias de los Materiales, Dr Moliner 50, 46100 Burjassot, Spain
                  \and A. Santangelo \at
                  Iniversit\"{a}t T\"{u}bingen
                  Institut f\"{u}r Astronomie und Astrophysik, Sand 1, 72076 T\"{u}bingen, Germany
                  \and
                  E. Churazov, R. A. Sunyaev \at
                  Max-Planck-Institut für Astrophysik, Karl-Schwarzschild-Str. 1, D-85748 Garching, Germany
                  \and S. Basso, M. Civitani, S. Covino, D. Spiga, G. Pareschi,  G. Tagliaferri, F. Tavecchio \at
                  INAF/Osservatorio Astronomico di Brera, Via E. Bianchi 46, 23807 Merate, Lc, Italy
                  \and R. Elsner, S. L. O'Dell, B. Ramsey, M. Weisskopf \at
                  NASA Marshall Space Flight Center, 320 Sparkman Drive NW, Huntsville, AL 35805-1912, USA
                  \and P. Attin\'{a} \at
                  Thales Alenia Space-Italia s.p.a., Strada Antica di Collegno 253, 10146 Turin, Italy
                  \and L. Baldini, M. M. Massai \at
                  Dipartimento di Fisica \\
                  Universita' di Pisa \& INFN-Pisa, Largo Pontecorvo, 3 56127, Pisa, Italy
                  \and R. Aloisio, E. Amato, P. Blasi, N. Bucciantini, M. Salvati \at
                  INAF/Osservatorio Astrofisico di Arcetri, Largo Enrico Fermi, 50125 Florence, Italy
                  \and L. Burderi \at
                  Universit\`{a} di Cagliari\\
                  Dipartimento di Fisica, SP Monserrato-Sestu km 0.7, 09042 Cagliari, Italy
                  \and V. Burwitz \at
                  Max-Planck-Institut f\"{u}r extraterrestrische Physik, D-85741 Garching \\
                  $\&$ Panter X-ray test facility, Gautinger Str. 45, D-82061 Neuried, Germany
                  \and P. Casella, G. Israel, L. Stella \at
                  INAF/Osservatorio Astrofisico di Roma, Via di Frascati, 33, 00040 Rome, Italy
                  \and R. M. Curado da Silva, J. Maia, T. Teixeira Dias \at
                  Dept. de Fisica, Coimbra Univ., Portugal, 3004-516 Coimbra, Portugal
                  \and G. Cusumano, S. Giarrusso, G. La Rosa, T. Mineo \at
                  INAF/IASF-Palermo, Via Ugo La Malfa 153, 90146 Palermo, Italy
                  \and M. Dadina, P. Grandi \at
                  INAF/IASF-Bologna, via Gobetti 101, 40129 Bologna, Italy
                  \and T. Di Salvo, G. Peres, F. Reale \at
                  DiFC, Universita' degli Studi di Palermo, via Archirafi 36, 90123 Palermo, Italy
                  \and M. Dovciak, V. Karas \at
                  Astronomical Institute\\
                  Academy of Sciences of the Czech Republic, Bocni II 1401, CZ-14131 Prague, Czech Republic
                  \and C. J. Eyles \at
                  University of Birmingham\\
                  School  of Physics and Astronomy, B152TT Birmingham, United Kingdom
                  \and H. Feng \at
                  Department of Engineering Physics \& Center for Astrophysics\\
                  Tsinghua University, 100084 Beijing, China
                  \and Andrew C. Fabian \at
                  University of Cambridge\\
                  Institute of Astronomy, Madingley Rd, CB3 0HA Cambridge, United Kingdom
                  \and R. W. Goosmann, N. Grosso, F.  Marin , D. Porquet \at
                  Observatoire Astronomique de Strasbourg, 11 rue de l'universit\'{e}, 67000 Strasbourg, France
                  \and P. Kaaret \at
                  Department of Physics and Astronomy\\
                  University of Iowa, IA 52242 Iowa City, USA
                  \and D. Lai \at
                  Space Sciences Bldg.\\
                  Cornell University, NY 14853 Ithaca, USA
                  \and A. Maggio \at
                  INAF/Osservatorio Astronomico di Palermo\\
                  Piazza del Parlamento 1, 90134 Palermo, Italy
                  \and P.-O. Petrucci \at
                  Institut de Plan\'{e}tologie et d$'$ Astrophysique de Grenoble (IPAG), UJF-Grenoble 1 /CNRS-INSU UMR 5274, Grenoble, F-38041, France
                  \and N. Rea \at
                  Institute of Space Sciences\\
                  CSIC-IEEC; Facultat de Ciencies, 08193, Barcelona, Spain
                  \and A. R\'{o}\.{z}a\'{n}ska \at
                  Polish Academy of Sciences\\
                  Nicolaus Copernicus Astronomical Centre, Bartycka 18, 00-716 Warsaw, Poland
                  \and P. Rudawy \at
                  Wroclaw University\\
                  Astronomical Inst, Kopernika 11, 51-622 Wroclaw, Poland
                  \and S. Sazonov \at
                  Space Research Institute, Russian Academy of Sciences,
                  Profsoyuznaya 84/32, 117997 Moscow, Russia\\
                  Moscow Institute of Physics and Technology, Institutsky per. 9,
                  141700 Dolgoprudny, Russia
                  \and E. Silver \at
                  Harvard-Smithsonian Center For Astrophysics, 60 Garden Street, MA 02138 Cambridge, USA
                  \and T. Tamagawa \at
                  RIKEN, 2-1 Hirosawa, Wako, Saitama 351-0198, Japan
                  \and M. van Adelsberg \at
                  Georgia Institute of Technology\\
                  Centre for Relativistic Astrophysics, 837 State Street, GA 30332 Atlanta, USA
                  \and K. Wu, S. Zane \at
                  University College London\\
                  Mullard Space Science Laboratory, Holmbury St Mary, RH5 6NT Dorking, United Kingdom
}

\date{Received: date / Accepted: date}

\maketitle

\begin{abstract}
X-ray polarimetry, sometimes alone, and sometimes coupled to spectral and temporal variability
measurements and to imaging, allows a wealth of physical phenomena in astrophysics to be studied.
X-ray polarimetry investigates the acceleration process, for example, including those typical
of magnetic reconnection in solar flares, but also emission in the strong magnetic fields
of neutron stars and white dwarfs. It detects scattering in asymmetric structures such as
accretion disks and columns, and in the so-called molecular torus and ionization cones.
In addition, it allows fundamental physics in regimes of gravity and of magnetic field
intensity not accessible to experiments on the Earth to be probed. Finally, models
that describe fundamental interactions (e.g. quantum gravity and
the extension of the Standard Model) can be tested.

We describe in this paper the X-ray Imaging Polarimetry Explorer
(XIPE), proposed in June 2012 to the first ESA call for a small
mission with a launch in 2017. The proposal was, unfortunately,
not selected.

To be compliant with this schedule, we designed the payload mostly with existing items.
The XIPE proposal takes advantage of the completed phase A of POLARIX for an ASI small
mission program that was cancelled, but
is different in many aspects: the detectors, the presence of a solar flare polarimeter and photometer and
the use of a light platform derived by a mass production for a cluster of satellites.
XIPE is composed of two out of the three existing JET-X telescopes with two Gas Pixel Detectors
(GPD) filled with a He-DME mixture at their focus.
Two additional GPDs filled with a 3-bar Ar-DME mixture always face the Sun to detect polarization from solar flares.

The Minimum Detectable Polarization of a 1 mCrab source reaches
14$\%$ in the 2$-$10~keV band in $10^{5}$~s for pointed
observations, and 0.6$\%$ for an X10 class solar flare in the
15$-$35~keV energy band. The imaging capability is 24~arcsec  Half
Energy Width (HEW) in a Field of View of 14.7~arcmin $\times$
14.7~arcmin. The spectral resolution is 20$\%$ at 6~keV and the
time resolution is 8~$\mu$s. The imaging capabilities of the JET-X
optics and of the GPD have been demonstrated by a recent
calibration campaign at PANTER X-ray test facility of the
Max-Planck-Institut f\"{u}r extraterrestrische Physik  (MPE,
Germany).

XIPE takes advantage of a low-earth equatorial orbit with Malindi as down-link station and of a
Mission Operation Center (MOC) at INPE (Brazil). The data policy is organized
with a Core Program that comprises three months of Science Verification Phase and $25\%$ of net observing
time in the following two years. A competitive Guest Observer program covers the remaining 75$\%$ of the net
observing time.

\end{abstract}

\section{Introduction}
\label{Intro} In 50 years of X-ray astronomy, instrumentation has
achieved fantastic advancements in imaging ({\it Chandra}), timing
({\it Rossi X-ray Timing Explorer}, {\it RXTE}) and spectroscopy
({\it Chandra} and {\it XMM-Newton}). No equivalent progress has
been achieved in X-ray polarimetry, despite the fact that the key
to uncover a number of scientific questions in fundamental physics
and the behavior of matter under extreme conditions is encoded
uniquely in this largely unexplored degree of freedom of
high-energy radiation. In spite of the lack of fresh data, solid
theoretical developments suggest that a wealth of important issues
on the physics of X-ray sources could be solved by measuring their
linear polarization.

At the beginning of X-ray astronomy, polarimeters were flown
aboard rockets (\cite{Angel1969, Novick1972}) and aboard the OSO-8
(\cite{Novick1975, Weisskopf1976}) and ARIEL-5 (\cite{Gowen1977})
satellites. The only positive detection was the polarization of
the Crab Nebula (\cite{Weisskopf1978}) and two significant upper
limits were obtained on Cyg X-1 (\cite{Weisskopf1977}) and Sco X-1
(\cite{Weisskopf1978b}), plus many other upper limits of modest
significance (\cite{Hughes1984}). The introduction of X-ray
optics, while producing a dramatic improvement in sensitivity,
removed the need to rotate the satellite. Therefore, polarimetry
based on the classical techniques, Bragg diffraction and Thomson
scattering (which require rotation), became seriously mismatched
with imaging and spectroscopy. As a result, no polarimeters were
included in major X-ray missions by NASA or ESA. Non-solar hard
X-ray polarimeters based on Compton effect resulted in a number of
balloon-borne narrow field experiments
(\cite{Gunji2010,McConnell2009,Pearce2012}) and in a polarimeter
for Gamma Ray Bursts (\cite{Yonetoku2011}) on-board the
solar-power sail demonstrator IKAROS.

In the last 10 years, with the development of sensors based on the
photoelectric effect (\cite{Costa2001}), polarimetry has been
again considered as a realistic option, either for large
telescopes with swappable instrumentation or for dedicated small
missions. An intense activity of theoretical modeling has started
again. A polarimetry mission, POLARIX (\cite{Costa2010}), was one
of two selected for flight after a phase A study following an ASI
AO issued in 2008 for a small mission. The program was
subsequently cancelled. Soon after, in 2009, NASA approved the
Gravity and Extreme Magnetism Small Explorer (GEMS,
\cite{Swank2010,Jahoda2010, Hill2012}), a small size satellite to
perform X-ray polarimetry, to be launched in 2014. Just at the end
of May 2012, NASA decided to discontinue GEMS for programmatic
reasons. A polarimetry mission based on instrumentation already
existing or of high technical readiness level became, therefore,
very timely.  The X-ray Imaging Polarimetry Explorer (XIPE)
fulfils these requirements. It is based on already existing items,
namely two out of three X-ray Mirror Modules built, tested and
calibrated for the JET-X project (\cite{Citterio1996, Wells1997})
and never flown (a fourth one is now operating well on the Swift
satellite \cite {Burrows2005}) and the GPDs studied for more than
10 years (\cite{Bellazzini2006, Bellazzini2007, Muleri2008,
Muleri2010, Soffitta2013}) and extensively tested for POLARIX and
for XEUS/IXO (X-ray Evolving Universe Spectroscopy
then evolved in International X-ray Observatory),
\cite{Bellazzini2010}). The photon by photon approach of the GPDs
and the wide, 14.7~arcmin $\times$~14.7 arcmin) are compatible
with a satellite of modest performance in terms of attitude control.

An extremely robust and relatively cheap bus used for
communication satellites, the Iridium NEXT, can harbor XIPE
without modification. XIPE can perform polarimetry of tens of
X-ray sources combined with imaging (24~arcseconds HEW
resolution), spectroscopy of continuum (20$\%$ @ 6~keV) and timing
(8~$\mu$s resolution). XIPE's unique results enable us to explore
the physics in extreme magnetic fields (in isolated or accreting
pulsars) and in extreme gravitational fields (in neutron stars and
black holes), to study the acceleration of particles in shocks in
supernova remnants and to study the disk and the onset of jets in
$\mu$quasars. A sample of extragalactic objects can also be
probed, especially Blazars. Due to the high readiness of the
technology, we are also proposing to perform polarimetry of solar
flares, which will provide a clue to understanding the physics of
magnetic reconnection.

XIPE opens a new window in high energy astrophysics and offers a large discovery space. Tests
of fundamental physics can be performed using the Universe as a laboratory with extreme phenomenology related
to General Relativity, the measure of the spin of black holes, or to QED, the detection of effects of
vacuum polarization in extreme magnetic fields. Last but not least, XIPE could search for the birefringence
predicted by Loop Quantum Gravity Theories or by theories of axion-like particles: one of the less exotic
but most elusive candidates for Dark Matter.

The breakthrough results promised by XIPE well fit the themes of ESA Cosmic Vision: 2.1 \emph{From the Sun to the edge
of the Solar System}; 3.1 \emph{Explore the limits of contemporary physics}; 3.3 \emph{Matter under extreme conditions} and 4.3
\emph{The evolving violent Universe}, and are far beyond what could be expected from a small mission.
This is only possible because we can use instrumentation of high performance and demonstrated maturity, including a
calibration of one GPD at the focus of an X-ray JET-X optics at the PANTER X-ray test facility. Such instrumentation,
in large part, already exists.

\section{Astrophysics with XIPE}
\label{sec:Astroph} XIPE, while proposing to install the same
X-ray optics of POLARIX (\cite{Costa2010}), takes advantage of a
new detector design (\cite{Muleri2012,Bellazzini2013}) involving a
larger body with improved control of the electric field and of the
space distribution of the residual background
(\cite{Soffitta2012}), and is also capable of measuring
polarization of radiation emitted by solar flares by means of a
GPD with an extended energy band up to 35~keV, and to make high
time resolution photometry. Moreover, the proposed platform is
directly derived from those of a  cluster of satellites for
telecommunication. The platform is lighter and includes the option
of transmission using the X-band.

The astrophysical goals of XIPE, except in the case of solar flares, follows the line of
what is already proposed for POLARIX. Hereafter, we summarize only the main theoretical
expectations, including new updates and the expected sensitivity in terms of
Minimum Detectable Polarization (MDP, see equation \ref{MDP}).

\subsection{Acceleration phenomena}
Acceleration phenomena in Supernova Remnants are believed to be
responsible for the production of the bulk of the cosmic rays
reaching the Earth, while jetted Active Galactic Nuclei (AGNs)
(\cite{Auger2007,Auger2012}) are a possible source of Ultra High
Energy Cosmic Rays. X-rays are emitted close to the region of the
maximum possible acceleration by means of the synchrotron
mechanism by electrons that then rapidly lose their energy. X-ray
polarimetry probing the environment close to the acceleration site
is therefore a powerful tool to investigate the acceleration
phenomena.

\begin{itemize}
    \item \textbf{Supernova Remnants} Supernova Remnants (SNRs) are believed to be the acceleration sites
                  of cosmic rays up to 10$^{15}$~eV. While the line emission makes it possible to determine the state
                  of ionization of its thermal plasma, the lack, or the weakness, of emission lines is generally
                  believed to be due to acceleration mechanisms responsible for the synchrotron emission or
                  non thermal bremsstrahlung.
                  Moreover, TeV emission from some SNRs supports the idea that in some regions, electrons are
                  energized at least up to TeV. Imaging polarimetry in this regard is useful to
                  localize the regions of shock acceleration and to measure the strength and
                  the orientation of the magnetic field at these emission sites (\cite{Vink2012}).
                  Probing the regions where thermal or non-thermal plasma is emitting in X-rays is
                  particularly important in small size SNRs like Cas A, Tycho and Kepler
                  (see e.g. \cite{Araya2010,Vink2012} and see fig.\ref{CASA_CRAB_ASIC}). {The high X-ray polarization
                  expected where the synchrotron process is prevalent (e.g. in the filaments usually located on the shell
                  boundaries) should be much reduced where the non-thermal emission is just a fraction of the thermal one.
                  However in the SNR 1006 radio polarization (\cite{Reynoso2013}) showed that where the
                  synchrotron is prevalent (e.g. in the two bright radio and X-ray lobes NE and SW) the measured degree of
                  radio polarization is just 17\% and this is probably due to a locally disordered magnetic field. Instead in
                  regions where non-thermal emission is not prevailing as in the SE rim the measured polarization is 60\%
                  possibly indicating a highly oriented magnetic field. Such considerations may be applied to Cas A for which
                  there is an indication of the presence of a tangential magnetic field at its outer edges (\cite{Gotthelf2001}).
                  We note that in this source (\cite{Bleeker2001} but see also \cite{Fabiani2013}) that in the spectral
                  region between 4 and 6 keV the power-low component is about 22.5 \% of the total emission while between
                  8 and 10 keV this component is 50 \%. Being the equivalent width of the iron line about 1 keV
                  the fraction of the power-law component between 6 and 8 keV is a non-negligible 19\%.
                  With one long (1 Ms) look of Cas A, the MDP is 1.6\% (4-6 keV), 3.5\% (6-8 keV) and 11.7\% (8-10 keV)
                  or 4.3\% 10.5 and 35\% in each of the 9 subregions that  we can think to divide Cas A. Giving the
                  estimated fraction of the power low component, polarization larger than 21\%  (4-6 keV), 55\% (6-8 keV) and  70\% (8-10 keV),
                  can be detected in each one of these subregions but interesting numbers could be obtained with just one energy
                  integration for a significative measurement. Based on the Einstein survey (\cite{Seward1990}), there are about ten SNRs with
                  a small ($<$ XIPE FoV) size having sufficient flux for X-ray polarimetry while the strategy for a space resolved
                  measurement can be implemented after having analyzed the observation of Cas A.
                  Oppositely large size SNRs ($>$ 30$'$) with a clear X-ray synchrotron spectrum in their rims are
                  SN 1006, RX J1713.7$-$3946, and RX J0852.0$-$4622. Clearly all these considerations
                  depend on how much the magnetic fields are ordered but this is precisely the scope of such measurements.}

    \item \textbf{Pulsar Wind Nebulae} Spatially resolved X-ray polarimetry allows
                  the magnetic field orientation in the torus,
                  in the jet and at various distances from the pulsar to be determined. This makes it
                  possible to evaluate the level of turbulence and
                  instabilities exploring the acceleration mechanism responsible
                  for the observed particle distribution (\cite{Shibata2003, Volpi2009}). XIPE reaches an MDP
                  of 2$\%$ in 5 $\times$ 5 angularly resolved regions of the Crab Nebula
                  in 10$^{5}$~s of observing time thanks to its imaging capability. The capability to
                  resolve the surrounding nebula makes polarimetry of the pulsar more straightforward,
                  allowing the emission model to be derived and compared for example with those studied in optical
                  band (\cite{Harding2005}).
                  A few additional PWNs will be accessible to XIPE for comparative measurements
                  (see fig. \ref{CASA_CRAB_ASIC}).

\begin{figure}[ht]
\centering
%% Use the relevant command to insert your figure file.
%% For example, with the graphicx package use
\includegraphics [scale=0.5] {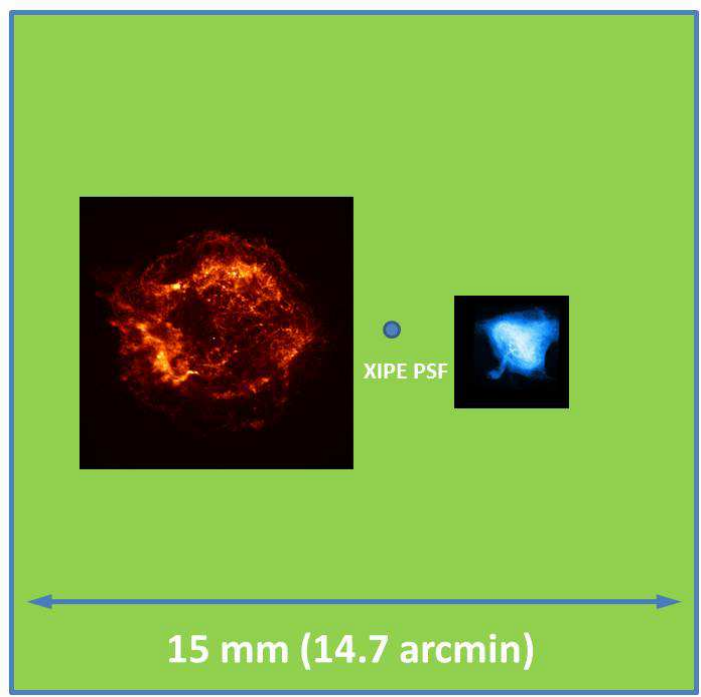}
\caption{The Chandra images of Cas A (left) and of the Crab Nebula
(right) within the sensitive area of XIPE together with its PSF
(Half Energy Width). The active area is 15 mm
$\times$ 15 mm or 14.7 arcmin $\times$ 14.7 arcmin.}
\label{CASA_CRAB_ASIC}       % Give a unique label
\end{figure}

    \item \textbf{Jets} The acceleration mechanisms in jets and the related X-ray emission, especially at large
    distances from the central massive objects, for both galactic and extragalactic sources, is
    a very much debated issue and X-ray polarimetry can, in both cases, help to resolve the matter.
    \begin{itemize}
        \item \emph{$\mu$QSOs} The multiwavelength behavior of about two dozen X-ray binaries with relativistic
        radio emitting spots, superluminal for a few of them, indicates that they are a scaled-down version of
        radio-loud galaxies, with consequently much shorter characteristic variability timescales. By means of
        spectro-polarimetry in X-rays and at other wavelengths of these very luminous objects, it is possible to shed light on jet formation
        and evolution and their relation with the
        accretion disk emission. GRS1915$+$105, Cyg X$-$1, Cyg X$-$3, and XTEJ1550$-$564 have flux between one hundred millicrabs and several
        crabs, allowing an MDP $<1\%$. They are also good candidates to search for General Relativity effects
        (see sec. \ref{GR_effects})

        \item \emph{Blazars $\&$ Radiogalaxies} In Blazars, multiwavelength polarimetry, including X-rays, would allow for
        disentangling the origin of the second characteristic emission peak in their spectral energy distribution, thanks to
        a determination of the polarization angle.
        This peak is due either to synchrotron-self Compton (same angle as that
        of the synchrotron peak (\cite{Celotti1994})) or to Inverse Compton (IC) of seed photons (different angle), presumably
        from the disk or from the broad-line regions. The degree of polarization of the IC peak allows the electron
        temperature (\cite{Poutanen1994}) in the jet to be estimated.  XIPE reaches an MDP of 3$\%$ for Mrk 421
        in 4 $\times$ $10^{5}$~s.

        In some radio-loud AGN, the jet component can be as bright as the disk component in the 2-10~keV energy band,
        as in 3C273 (\cite{Grandi2004, Grandi2007}). In this case, because the jet component is harder, a rotation of the polarization angle is expected.

    \end{itemize}

\item \textbf{Magnetic reconnection} Magnetic reconnection and
subsequent acceleration of charged particles in the corona are at
the base of the production of solar flares (\cite{Brown1971}).
Actually, the Sun, providing a strong signal due to its closeness,
acts as a Rosetta stone, clearly showing phenomena similar to
those which may happen in objects significantly fainter and much
farther away. Above 20~keV, see fig.\ref{fig:flare_loops} (left),
the emission from solar flares (Hard X-Ray, HXR) is mostly
dominated by the non-thermal bremsstrahlung generated by high
energy electrons impinging down on the chromosphere with a
polarization degree as high as 40$\%$ at 20~keV
(\cite{Zharkova2010}) (see fig. \ref{fig:Zharkova} left
and right, and its caption). Below 10~keV, the
emission is mostly thermal due to plasma heating in the
reconnection site and in the flaring loop filled with evaporated
chromospheric plasma (see fig. \ref{fig:flare_loops}, right).
X-ray lines are present up to 7~keV
(\cite{Peres1987,Doschek2002}). The thermal
component is also expected to be polarized, although at a lower
level than the non-thermal one, due to possible
anisotropies in the electron distribution function
(\cite{Emslie1980}). Back-scattering further modifies the
spectrum, especially at higher energies
(\cite{Bai1978,Jeffrey2011}). X-ray polarimetry of the HXR offers
the possibility to make a diagnostic of the level of the
anisotropy of the electron beams, and of the magnetic field
configuration, and to study the acceleration mechanism in the
solar corona. Both RHESSI, with its spectrometer not designed to
be a polarimeter (\cite{Suarez-Garcia2006}), and the Thomson
scattering polarimeter on-board Coronas-F (\cite{Zhitnik2006})
attempted to measure the X-ray polarization from solar flares with
only low significative results or large upper limits due to the
high energy threshold of the former and to the high background/low
efficiency of the latter. Moreover, future missions such as Solar
Orbiter are not sensitive to polarization. XIPE performs
polarimetry of radiation emitted by solar flares in the
15$-$35~keV energy band, reaching, for the two detector array
configuration, an MDP of 0.6$\%$ for an X10 class flare and
6.6$\%$ for an M5.2 class flare. We estimated the sensitivity to
different classes of solar flares using the spectra and the
lightcurves in \cite{SaintHilaire2008} in equation \ref{MDP}. We
also evaluated the expected number of flares for each class,
depending on the solar activity. Based on the forecast of the sun
spots\footnote{{http://www.swpc.noaa.gov/ftpdir/weekly/Predict.txt},2012/04/04}
we estimated that from July 2017 to June 2019 about two dozen
flares are expected to be observed between class X10 and class M5.
Being close to the solar minimum, the probability of an X flare
is, however, small. XIPE solar polarimeter operates
in an energy range where the spectrum of the flares is dominated
by non-thermal bremsstrahlung but the flux expected is large. By
measuring their polarization with good accuracy for a number of
flares in different positions on the solar disk, it will be able
to constrain different models (see fig. \ref{fig:Zharkova} right
and its caption) with much higher precision with respect to the
data available today.

XIPE will also monitor the X-ray variability of the Sun between
1.2 and 15~keV with a small dedicated photometer with good energy
resolution to determine the coronal average temperature and the
related thermodynamic characteristics for either non-active or
flaring corona. This also is particularly important for space
weather studies.
\end{itemize}

\begin{figure}[htpb]
\centering
\subfigure[\label{fig:flare_loops}]{\includegraphics[scale=0.40]{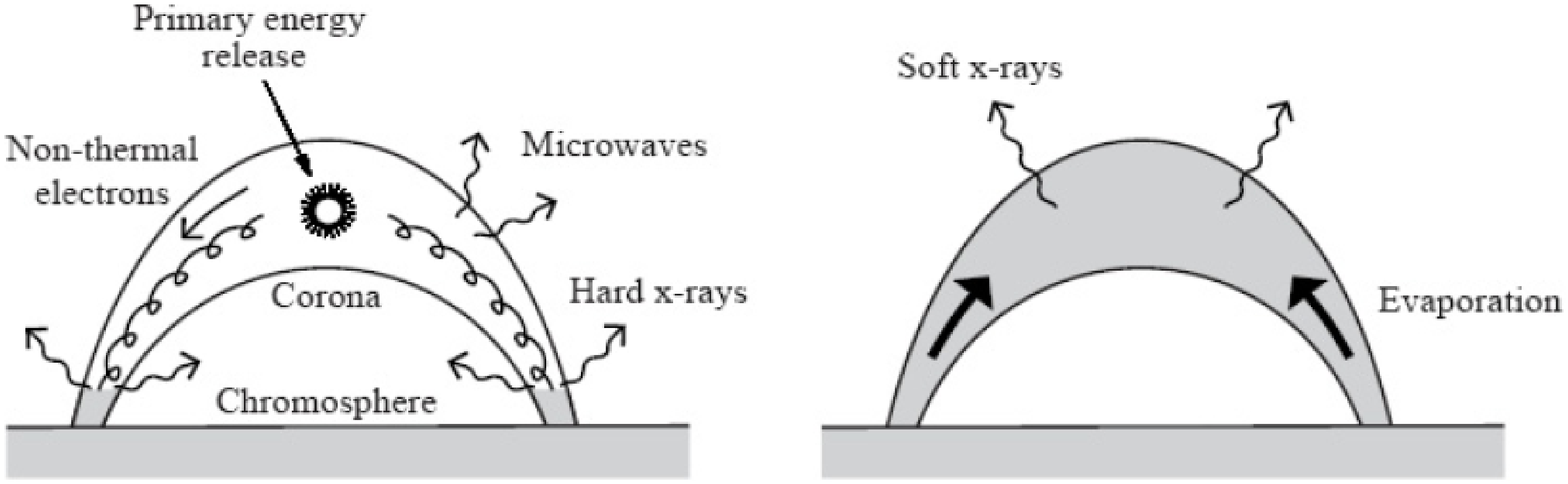}}
\hspace{1cm}
\subfigure[\label{fig:Zharkova}]{\includegraphics[scale=0.20]{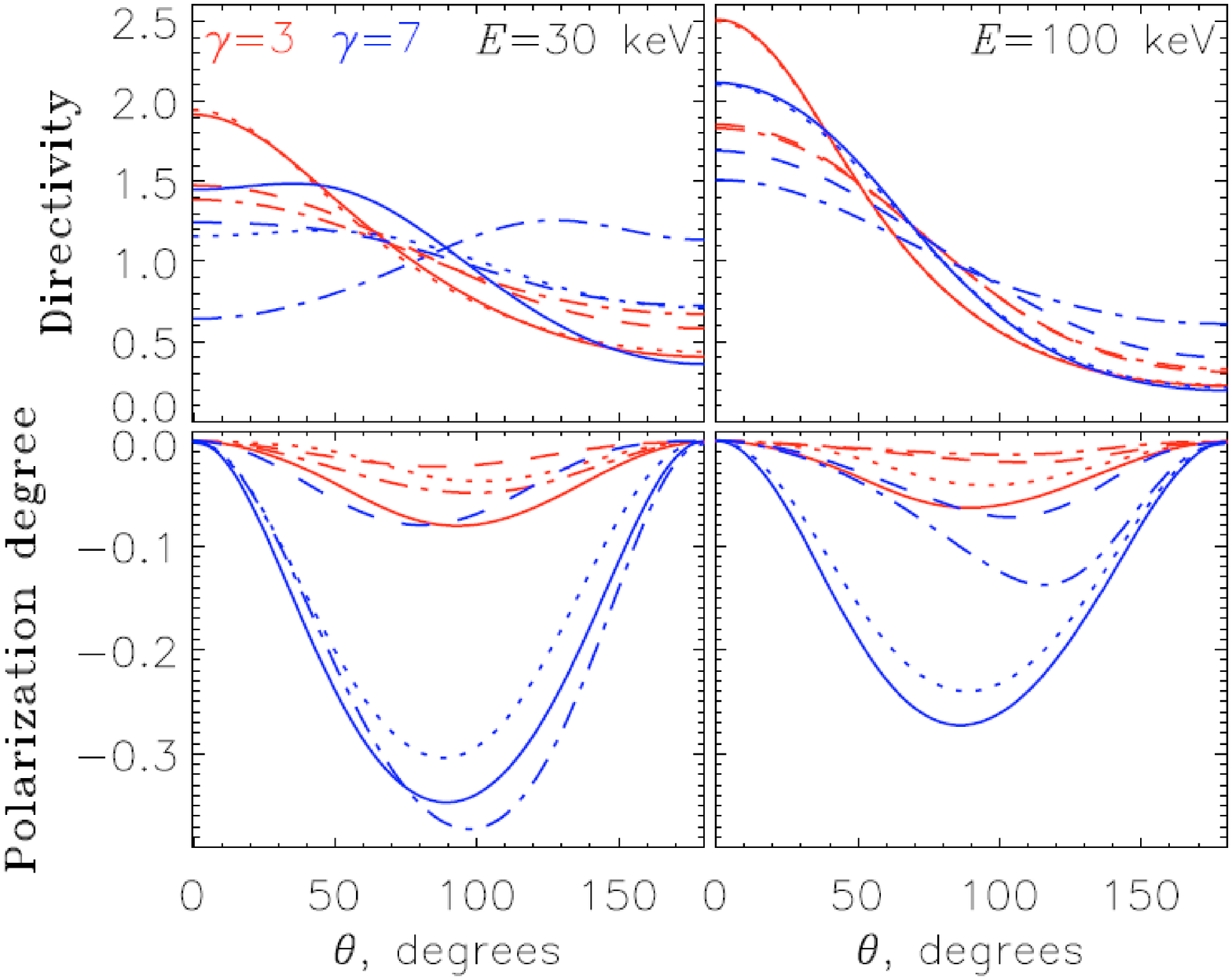}
\includegraphics[scale=4.0]{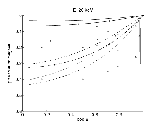}}
\caption{ ({\bf a}). Sketch of the flaring loop in hard X-rays and
soft X-rays (figure from \cite{Priest2002}) ({\bf b (left)}).
Directivity and polarization degree resulting from
the integration over all the coronal magnetic tube. The
directivity is the ratio between the intensity at a given angle
and the average intensity (unit value means isotropic emission).
The model explores two initial different power-law
indices of the particles that are accelerated : $\gamma = 3$
(\textcolor[rgb]{0.98,0.00,0.00}{red} lines) and $\gamma = 7$
(\textcolor[rgb]{0.00,0.00,1.00}{blue} lines). Different  lines
show different models for the simulation: solid line: pure
collisions(C), dashed line: collisions and converging magnetic
field (C$+$B), dotted line: collisions and return current (C$+$E),
dash-dotted line: all factors are taken into account (C$+$E$+$B).
$\theta$ is the radiation propagation direction that is the angle
between the normal to the sun where the injection and the downward
beaming occur and the observer. A higher polarization is expected
when $\theta$ is $90^\circ$ therefore when the flare is located on
the limb. ({\bf b (right)}). Comparison of the polarization
expected from different models with available data at 20~keV as a
function of $\Xi$ that is the position angle in the solar disk
(cos($\Xi$) = 1 is the disk center and 0 is the limb) for an
electron beam with wide angle dispersion ($\Delta\mu = 0.2$ where
$\mu$ is the cosine of the pitch angle of the precipitating
electrons). The electrons' energy flux is 10$^{10}$ erg cm$^{-2}$
s$^{-1}$. For $\gamma$ $=$ 3 the following models are shown: solid
line C$+$E model, solid line with crosses: C$+$E$+$B model. For
$\gamma$ $=$ 7 the models are : dashed line C$+$E model, dashed
line with crosses C$+$E$+$B model. The cases of a more collimated
electron beam ($\Delta\mu$ $=$ 0.09 with $\gamma$ $=$ 7 and C$+$E
model) are also shown. The initial electron energy fluxes are
respectively 10$^{10}$ erg cm$^{-2}$ s$^{-1}$) (dot-dashed lines)
and 10$^{12}$ erg cm$^{-2}$ (dotted lines). The data are from
\cite{Tindo1970,Tindo1972a,Tindo1972b} (at 15~keV, diamonds) and
from \cite{Tramiel1984} (16$-$21~keV, triangles). The figures are
from \cite{Zharkova2010}} \label{Solarflares}
\end{figure}

\subsection{Emission in strong magnetic fields}
The presence of an intense magnetic field affects the propagation of the two X-ray
polarization modes in a plasma.  Moreover, it channels the plasma along the field lines, causing an
asphericity in its distribution. Both phenomena produce radiation which is observed as anisotropic and polarized.

\begin{itemize}
    \item \textbf{Accreting White Dwarfs} In White Dwarfs (WDs) with a strong magnetic field, X-ray polarization derives from
the scattering on the WD surface, from its accreting column and,
when present, from the (truncated) disk. The polarization signal
is periodic and energy dependent, with values ranging from 4$\%$
to 8$\%$ (\cite{Matt2004, McNamara2008}). XIPE can search for
phase-dependent X-ray polarization in the brightest objects. For
AM Her, XIPE reaches an MDP of 6$\%$ in each of ten phase bins
with $10^{6}$~s of observation.

    \item \textbf{Millisecond X-ray pulsars} Accretion is responsible for the spin-up of neutron stars up to the maximum possible
rotation speed. Compton scattering in the accretion shock, which
is localized and not extended as shown by pulsation, polarizes the
radiation at higher energies (\cite{Viironen2004}). Phase resolved
X-ray polarimetry allows for testing this model and, possibly, for
discriminating an alternative scenario where the scattering is
from the accretion disk (\cite{Sazonov2001}). It also provides the
geometrical parameters, such as the orbital and magnetic
inclination, which are usually free parameters in the evaluation
of the mass and the radius of the neutron star. At
the present time 14 accreting millisecond X-ray pulsars (AMXP) are
known and they are very faint in quiescence. They, however, can
serendipitously outburst for several days with fluxes exceeding
tens of milliCrabs and more, showing their kilo-Hz pulsation. At a
flux of 10 mCrab, rather low for this kind of sources, and
integrating for 10$^6$ s, e.g. SAX J1808$-$4$-$3658 reaches an MDP
of 3$\%$ in 5 phase bins that is sufficient for modeling the
source.
    \item \textbf{Accreting X-ray pulsars} In accreting X-ray pulsars, the large magnetic field
    $(10^{12}-10^{13}$~Gauss) derived from the observed cyclotron lines creates birefringence
effects with an energy and phase dependent polarization signature
(\cite{Meszaros1988}). Phase resolved X-ray polarimetry allows for
determining the geometry of the accretion (fan or pencil beam),
the position of the rotation axis in the sky and the angle between
its position and the magnetic dipole. Many X-ray pulsars can be
observed by XIPE with sufficient sensitivity. For example, an
observation of Her X-1 is characterized by an MDP of 3.5$\%$ in 10
independent phase bins.
\end{itemize}

\subsection{Scattering in aspherical situations}

\begin{itemize}

    \item \textbf{X-ray binaries}
The 2$-$10~keV spectrum of accretion-disc-fed X-ray binaries in
the hard state is probably mostly due to Comptonization by a hot
corona. The aspherical geometry produces polarized X-rays
(\cite{Haardt1993, Poutanen1993, Poutanen1996b}) and the
polarization degree places constraints on the unknown geometry of
the hot corona (\cite{Schnittman2010}). Above 7~keV, the Compton
reflection of the primary emission from the disk is also expected
to be polarized with a polarization degree that depends on the
disk inclination and on the anisotropy of the intrinsic emission
(\cite{Matt1989, Poutanen1996}).

\item \textbf{Radio-quiet AGNs} In radio-quiet AGNs, the
Comptonization in the corona and the Compton reflection from the
disk always dominate with respect to the disk emission mostly
irradiating in UV or in soft X-ray, and the same considerations as
above apply (\cite{Schnittman2010}). A XIPE observation of IC4329A
of 3$\times 10^{5}$~s yields at an MDP of 3.6$\%$. In addition to
the above reflection environments, scattering can occur in AGNs on
the so-called molecular torus, whose geometry is still largely
unknown, and on the ionization cones when present (see fig.
\ref{Goosman}). X-ray polarimetry (see fig.
\ref{fig:GoosmanPolarization}) can shed light on the connection
between these two regions as well as on the true torus geometry
(\cite{Goosmann2011}). In the case of NGC 1068, an MDP of 4.2$\%$
can be reached with an observation lasting 5$\times$$10^{5}$~s.
With an observing time of 10$^{6}$ s the 3-$\sigma$ measurement is at level of 5\%
(2-4~keV) and 6.9$\%$ (4-10~keV). With this sensitivity it is
possible to disentangle most of the models and to provide an
additional 1-$\sigma$ error on the angle of 9.5$^\circ$ (see
paragraph \ref{sciencereq}) that is sufficient to hint the
possible rotation of 60$^\circ$. A multiwavelength polarization
campaign can allow for a deeper investigation of the geometries of
the scattering regions (\cite{Goosmann2011, Marin2012c}) in AGNs.

\begin{figure}[htpb]
\centering
\subfigure[\label{fig:GoosmanCones}]{\includegraphics[scale=0.15]{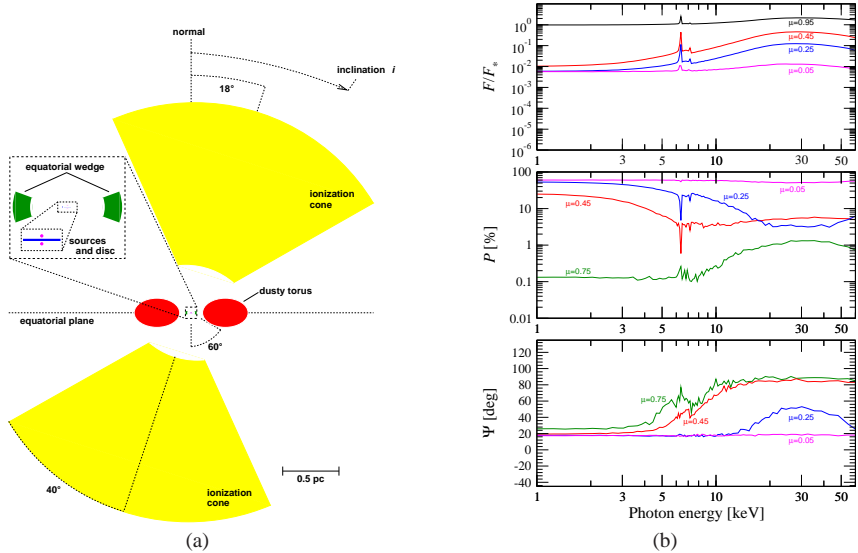}}
\hspace{1cm}
\subfigure[\label{fig:GoosmanPolarization}]{\includegraphics[scale=0.25]{Goosman_2011_fig5a.eps}}
\caption{({\bf a}). Sketch of the scattering environment around
NGC 1068. ({\bf b}). Intensity, level of polarization and
polarization angle as a function of energy for the model of fig.
\ref{fig:GoosmanCones}. Here the column density of the torus is
$N_H=10^{27}$~cm$^{-2}$ and the optical depth of the scattering
cones is $\tau_{cone}$ = 0.3.
$\mu$ = cos(i) where
i is the viewing direction measured with respect to the disk and
torus symmetry axis; F$_{\star}$ is the total flux of the primary
source, emitted into the same viewing direction. Both figures
are from \cite{Goosmann2011}.} \label{Goosman}
\end{figure}

\item \textbf{X-ray reflection nebulae} There are a few molecular
clouds in the Galactic Center region whose X-ray spectra are well
reproduced by a pure Compton Reflection component, indicating that
such clouds are reflecting the X-ray radiation produced by a
source outside the cloud. The most famous example is Sgr B2, but
more recently, the X-ray emission from the Sgr C complex was
additionally proposed to have the same origin
(\cite{Murakami2001_SGRC}). The puzzle here is that there is no
X-ray source bright enough in the surroundings. It has been
proposed, therefore, that these clouds are reflecting past
emission from the central black hole (\cite{Sunyaev1993,
Koyama1996}), which should have undergone a phase of strong
activity about three hundred years ago. If the emission from the
nebulae is indeed due to scattering, it should be very highly
polarized (\cite{Churazov2002}), with a direction of polarization
normal to the scattering plane, and therefore to the line
connecting the cloud to the illuminating source. The detection of
polarized X-ray emission from one or more of these clouds would
place a strong limit on the position of the source which
illuminated them in the past and, if the polarization plane is
indeed perpendicular to the direction towards Sgr A*, it will be
proved that not many years ago the Galaxy was a low luminosity
AGN. In addition, measurements of the polarization degree will
provide unique information on the position of the clouds with
respect to Sgr A* along our line of sight. The flux from Sgr B2 is
evolving with time. It is currently decreasing
(\cite{Koyama2008}), probably reflecting the evolution of the
illuminating source flux in the past. Other reflecting nebulae are
present around the central black hole, which are also varying with
time, e.g. Sgr C (\cite{Murakami2001_SGRC, Muno2007}), and the
brightest of them when XIPE will be in orbit will of course be
chosen for observation. Although it is not possible to estimate
the flux of Sgr B2 when XIPE will be in orbit, assuming the flux
measured by {\it BeppoSAX} (\cite{Sidoli2001}) and a polarization
of 40$\%$, the precision with which the polarization angle can be
measured in 2$\times$$10^{6}$~s is 3.5$^\circ$ (1-$\sigma$), good
enough to set tight constraints on the origin of the illuminating
radiation. One question is the background rate expected when
observing this very faint extended X-ray source. This question is
answered in section \ref{background}. Here we just write that the
background is still about 30 times smaller with respect to the
expected source rate as based on estimates from \cite{Bunner1978}.
The image of Sgr B2  on the detector active  area is shown in the
collage of fig. \ref{SGRB2_ASIC}.

\begin{figure}[ht]
\centering
%% Use the relevant command to insert your figure file.
%% For example, with the graphicx package use
\includegraphics [scale=7.5] {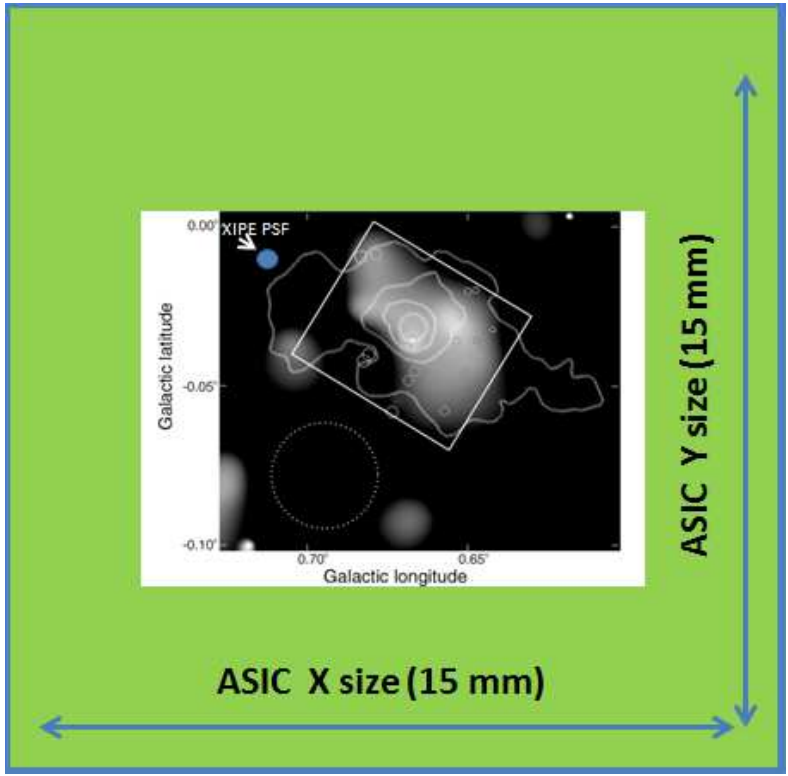}
\caption{Image of Sgr B2 on the GPD active area as in
\cite{Muralkami2001_SGRB2}. For XIPE, we assume half of this
region where we evaluated the background. The green area
represents the total active area of the X-ray polarimeter. The
galactic coordinates shown are in degrees. The active area is 15
mm $\times$ 15~mm or 14.7~arcmin $\times$ 14.7~arcmin.
The XIPE PSF is the Half Energy Width.}
\label{SGRB2_ASIC}       % Give a unique label
\end{figure}
\end{itemize}

\section{Fundamental physics with XIPE}
High Energy Astrophysics makes accessible natural laboratories of fundamental physics,
providing tests of physical theories which would otherwise be impossible.
X-ray polarimetry is a sensitive probe because distinctive signatures on the
degree and angle of polarization are expected during photon transfer in strong
gravitational or magnetic fields. Energy dependent rotations of the polarization angle
and variations of the polarization degree from distant sources may
reveal Quantum Gravity effects and allow for axion$-$like particle searches.

\subsection{QED in strong magnetic fields}
Emission from a neutron star (NS) surface is expected to be
polarized because the opacity of the atmosphere to photons
polarized perpendicular to the magnetic field (X-mode) is smaller
than that for parallel polarized radiation (O-mode)
(\cite{Pavlov1978}). The radiation in the X-mode can escape from
inner (and, therefore, hotter and brighter) layers of the NS
atmosphere. The expected degree of polarization induced by this
effect is not large, at level of 5$-$25$\%$ (\cite{Pavlov2000}),
because of the different magnetic field orientations at the
emission sites. Although this basic result is commonly accepted,
the detailed photon transfer across the atmosphere is strongly
affected by quantum-electrodynamics (QED) effects because a
magnetic field in excess of 10$^{14}$~Gauss for magnetars
polarizes the vacuum (\cite{Meszaros1992}). As discussed in
\ref{vacuum_resonance} and in \ref{birefringence}, QED produces
three detectable effects on X-ray polarization. One effect is on
the energy dependence of X-ray polarization degree and angle, due
to the presence of a vacuum resonance. The other two effects are
the enhancement of the maximum degree of polarization and the
observable lag of the polarization angle between optical light and
X-rays, both due to birefringence. The signatures on the spectrum
(softening of the hard tail and reduction of the equivalent width
of the proton-cyclotron line) can be less evident and, at the same
time, the modelling is affected by the degeneracy on the various
parameters.  Polarization measurements can be used to disentangle
these degeneracies (\cite{vanAdelsberg2009}).

\subsubsection{The effect of the vacuum resonance}
\label{vacuum_resonance} 
\begin{figure}[ht]
\centering
%% Use the relevant command to insert your figure file.
%% For example, with the graphicx package use
\includegraphics [scale=0.4] {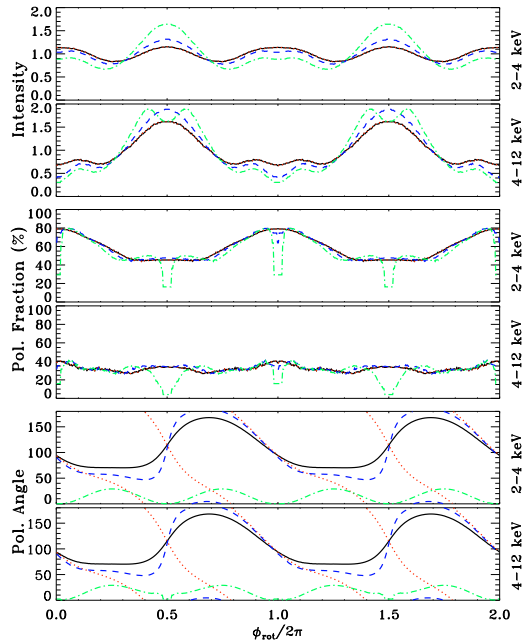}
\caption{This figure (from \cite{Fernandez2011}) is
the result of a Monte Carlo simulation of the phase-resolved
emission observed from a magnetar with an arbitrary flux modelled
as t12g18c90Eu ($\Delta\phi$NS(rad)=1.2,
$\gamma$max = 1.8, seed pol=E$-$mode, $\Theta_{cap}$($^\circ$)=90,
where $\Delta\phi$ is the twist angle of its magnetosphere,
$\gamma$max is the maximum Lorentz factor of the electrons in the
magnetosphere, $\Theta_{cap}$ is the angular size of the emitting
polar cap). From top to bottom are shown the intensity, the
polarization fraction and the polarization angle as a function of
rotational phase (two full periods shown) in energy bands
($2-$4~keV and 4$-$12~keV). The pair of angles ($\theta_{rot}$
the angle between the magnetic axis and the rotation
axis, $\theta_{los}$ the line of sight from the
rotation axis) of ($45^\circ$, $70^\circ$) is represented by a
solid/black line, ($70^\circ$, $45^\circ$) is
dotted/\textcolor[rgb]{0.98,0.00,0.00}{red}, ($60^\circ$,
$70^\circ$) is dashed/\textcolor[rgb]{0.00,0.00,1.00}{blue} and
($90^\circ$, $90^\circ$) is dot-dashed/\textcolor[rgb]
{0.00,1.00,0.00}{green}. It can be seen that the polarizations in
the two energy bands are always in phase.}
\label{Fernandez_fig10}       % Give a unique label
\end{figure}

Detailed calculations
(\cite{vanAdelsberg2006,Fernandez2011}) have shown that QED
effects have a peculiar signatures on the polarization of the
radiation detected by a distant observer, while they have a less
obvious impact on spectral parameters. A resonance, occurring when
the contribution to dielectric tensor of the plasma and of the
vacuum compensate each other, should produce a transition between
the two photon modes analogous to the Mikheyev-Smirnov-Wolfenstein
mechanism for neutrino oscillation (\cite{Lai2003}). The
probability of a transition between the photon modes depends on
the geometry, magnetic field, and properties of the medium,
increasing monotonically with the energy of the photon. For the
magnetic field regime B~$<$ 7$\times$~10$^{13}$~Gauss, the
resonance typically lies outside the photospheres of the two
photon modes. Since surface emission is usually dominated by the
X-mode, at energies $\gtrsim$ a few~keV, X-mode photons are
converted to O-mode photons after decoupling from the atmosphere,
leading to a rapid 90$^\circ$ shift in the angle of polarization.
For stronger, magnetar-strength magnetic fields, the resonance
occurs inside the O-mode photosphere; thus, the rapid rotation in
the plane of polarization does not occur (see fig.
\ref{Fernandez_fig10}). Detecting the 90$^\circ$ rotation in
normal pulsars but not magnetars would independently confirm the
presence of super-strong magnetic fields in magnetars. Such
rotation can be excluded for example in the case of SGR 1806-20 a
bright (not the brightest) magnetar.
In $10^{6}$ seconds
we can get a 3-$\sigma$ measurement at level of 23.6\% (2$-$4~keV)
and 21.3\% in (4$-$10~keV) for each of the 5 phase bins. This
level of polarization, well below the expected value from the
model of \cite{Fernandez2011} showed in the figure
\ref{Fernandez_fig10}, implies a 1-$\sigma$ error in angle of
9.5$^\circ$, a precision that is sufficient to exclude the
90$^\circ$ rotation in the data.

\subsubsection{The effect of birefringence} \label{birefringence}
X-ray polarimetry of NS emission provides an opportunity to
observe another QED effect: vacuum birefringence induced by a
strong magnetic field. This effect was predicted nearly 70 years
ago (\cite{Heisemberg1936, Weisskopf1936}) but still needs to be
verified experimentally. If the vacuum birefringence is present,
the indices of refraction of the two linear polarization modes
differ from each other. Vacuum polarization produced by a NS
magnetic field is indeed sufficient to decouple the polarization
modes, so that the direction of polarization follows the direction
of the local magnetic field \cite {Heyl2000}. When modes are
coupled again at a distance which is large with respect to the
radius of the NS, the local magnetic field is almost parallel and
therefore photons coming from different regions of the NS surface
add coherently. This produces a 5--7 times larger polarization
degree in the NS phase averaged signal (\cite{Heyl2003}) than in
estimates where birefringence is not taken into account
(\cite{Pavlov2000}).

In addition, the NS and its magnetosphere rotate and, since the
modes of lower energy radiation couple first, another observable
prediction of the presence of vacuum birefringence effects is that
the angle of polarization at low energy should lag behind higher
energy photons (\cite{Heyl2000}). For the Crab pulsar,  the lag
between X-rays and optical emission should be about 10 degrees, as
derived in the Deutsch model (\cite{Deutsch1955}) in
\cite{Heyl2000}.

\subsection{General Relativity in extreme gravity fields}
\label{GR_effects} 
\begin{figure}[htpb]
\centering
\subfigure[\label{fig:Poldeg}]{\includegraphics[scale=0.55]{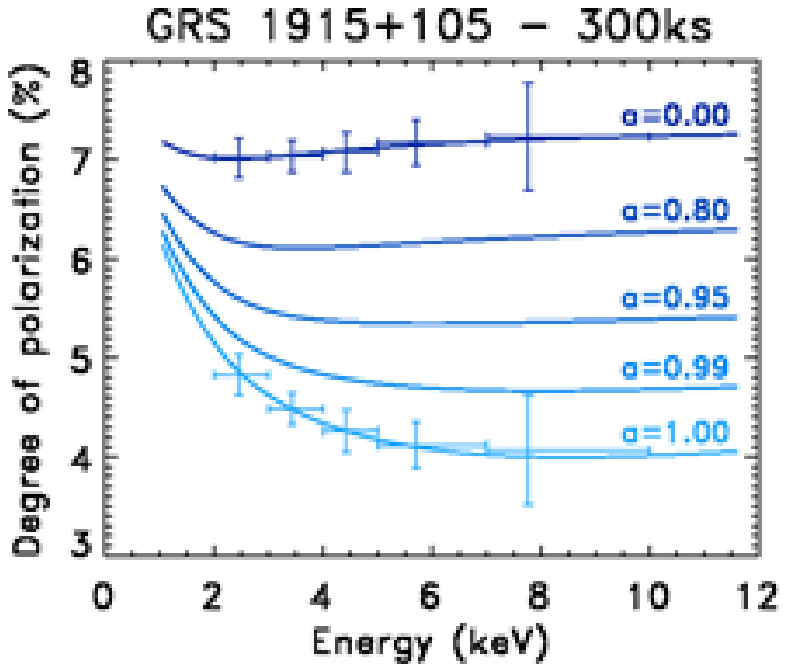}}
\hspace{1cm}
\subfigure[\label{fig:Polrot}]{\includegraphics[scale=0.55]{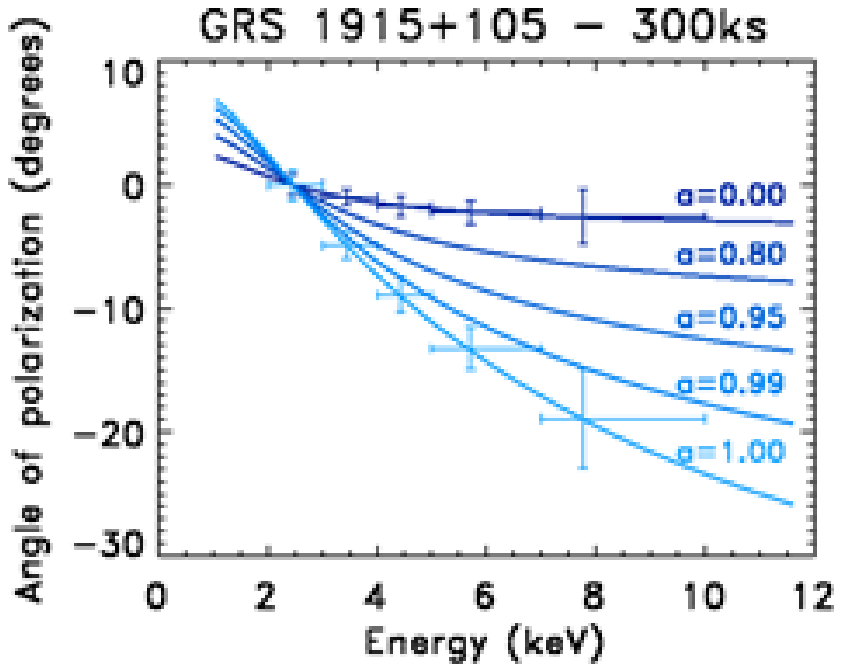}}
\caption{({\bf a}). Expected variation of the polarization degree
with energy in GRS1915+105 simulating an observation of 300~ksec
The model is from \cite{Dovciak2008} while the errors are evaluated
for the case of an observation with XIPE). ({\bf b}).
Polarization angle rotation with energy in the same observation.} \label{GRS1915}
\end{figure}

The emission from the accretion disk is the
brightest component of the X-ray spectrum from Galactic Black
Holes when they are in a high state. The strong gravitational
field in the innermost region of the disk, that is responsible for
the X-ray emission, causes a rotation in the polarization angle
larger than for a rotating black-hole and for higher energies
(\cite{Stark1977, Connors1980, Dovciak2008, Li2009,
Schnittman2010, Krawczynski2012}). The measurement of the rotation
of the polarization angle (and degree) with energy, and therefore
of the spin of the black-hole, allows for testing General
Relativity in extreme gravity fields. The best but not the only
source to search for this effect is GRS1915+105 (see fig.
\ref{GRS1915}), a bright $\mu$QSO whose 2-10~keV emission is, when
in high state, dominated by thermal emission. Moreover, the source
is highly inclined (70$^\circ$ in \cite{Mirabel1994}), and
therefore the polarization degree is expected to be high. Other
less inclined sources may show lower polarization levels, which
could still, however, be easily detected in several other bright
objects. In addition, about 4 transient BH binaries are expected
to have a large enough flux to be measured during 2 years of
operation and they are good sources to search for GR effects.

\begin{figure}[h]
\centering
%% Use the relevant command to insert your figure file.
%% For example, with the graphicx package use
\vspace{0.1in}
\includegraphics [scale=0.25, angle=-90] {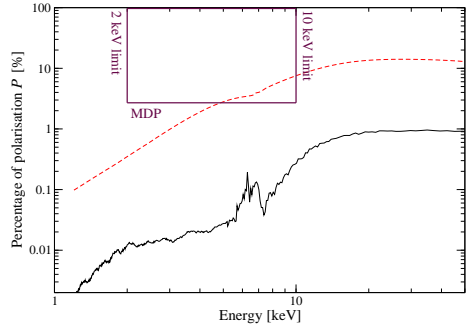}
\caption{MDP of the two scenarios for a 1 Ms observation of
MCG$-$6$-$30$-$15. The solid curve represents clumpy absorption,
while the red dashed curve is relativistic reflection induced by a
Kerr super massive black hole with spin parameter $a=1$.
With an observation of 2 Ms a
polarization larger than 3\% (MDP) can be detected between 4 and
10 keV to be compared with a higher (3.6\%) expected average
polarization degree, after spectral convolution in the same energy
band. The figure is from \cite{Marin2012b}}
\label{Marin_fig3}       % Give a unique label
\end{figure}

In AGNs, the thermal emission from the disk peaks in UV and
therefore lies outside the energy band of XIPE. Strong Gravity
effects manifest themselves with the temporal variation of the
polarization angle from reflected X-rays (\cite{Dovciak2011}) from
a primary source with changing height from the accretion disk (see
 \cite{Miniutti2004}), as in the case of MCG$-$6$-$30$-$15. In this
source, an MDP of about 4$\%$ can be reached in 300~ks. A long
look (1~Ms or more) at this source may provide a first test of the
model. Alternatively, there is at least one galactic black hole
candidate (XTE J1650$-$500) which is thought to behave like
MCG$-$6$-$30$-$15 on a smaller scale (\cite{Rossi2005, Reis2013}).
Thanks to a factor of 100 higher flux, this source would allow for
deeper studies of this phenomenon. An alternative model explains
the observed relativistic iron line from MCG$-$6$-$30$-$15 as a
non-relativistic feature arising from partial covering
 (\cite{Miller2009}). Partial absorption in a clumpy outflow
intercepting the line of sight generally induces low-polarized
forward scattering and always produces a polarization position
angle that is constant in energy. In the reflection case, on the
other hand, the polarization is larger and its position angle
varies systematically with energy.
A larger
polarization is expected at higher energies and a long-look
observation of 2~Ms can detect polarization above 3\% (4$-$10~keV)
as expected by the the reflection model (\cite{Marin2012b}) only,
and thus strongly favor one of the two interpretations, see fig.
\ref{Marin_fig3}. Furthermore, (\cite{Horak2006}) pointed out
two specific effects of general relativity that can be revealed in
linear polarization from light scattered by relativistic jets that
are expected due to indirect photons passing in the immediate
vicinity of a black hole.

\subsection{Quantum Gravity}
The identification of a good candidate observational test for
studying Quantum Gravity in its different forms (loop, string,
non-commutative space-times) is presently still a challenge
(\cite{Amelino2004}). Polarimetry is one of the few possible
probes (\cite{Gambini1999}) of loop Quantum Gravity. At the
quantum scale, birefringence would be responsible for a rotation
of the polarization angle along the photon path. Such rotation, in
the linear case, is proportional to the source distance and to the
square of the energy by means of an dimensionless factor $\eta$
(\cite{Mitrofanov2003}). The scale of breakdown of the usual
dispersion relation is $\eta$ $=$ 1, the Plank scale (see for
example \cite{Fan2007}). Detecting a non-vanishing linear
polarization from distant sources allows upper stringent limits to
be placed on $\eta$, and cases other than linear to be excluded.
Upper limits based on these observational tests already rely on
the UV polarization of a radio galaxy (\cite{Gleiser2001}) and on
X-ray polarization of the Crab Nebula measured by  OSO$-$8 as in
\cite{Kaaret2004} ($\eta$~$<$~10$^{-4}$). A more stringent upper
limit comes from the UV/optical polarization of Gamma-ray burst
afterglows as in \cite{Fan2007} ($\eta$~$<$~10$^{-7}$). In hard
X-rays, upper limits from INTEGRAL data (see, for example,
\cite{Maccione2008,Stecker2011,Laurent2011}) are instead based on
results derived from the prompt emission of Gamma Ray Bursts
 (\cite{Kalemci2007, McGlynn2007}) and from the Crab  emission (\cite{Dean2008}).
Such low-significance results are still debated. Sometimes,
different instruments on-board are themselves in contradiction. We
stress that the detectors were not primarily designed as
polarimeters. Such instruments were never calibrated on-ground or
in-orbit for this purpose. In this regard \cite{Toma2012} noted
that the data in the measured modulation curves in
\cite{McGlynn2007} are not distributed with Poissonian statistics
(while this is assumed in the evaluation of the error contours for
example in \cite{Dean2008}). Recently, by using the measurement (P
= $84^{+16}_{-28}$, 3.3~$\sigma$ significance) for GRB $110721A$,
with a known redshift of 0.382 (see \cite{Yonetoku2012} and
reference therein) by the GAP GRB Compton polarimeter aboard
IKARUS, an upper limit $\eta$~$<$~$10^{-15}$ was evaluated in
\cite{Toma2012}. Looking to different sources at different
distances the relation between the latter and the rotation of the
polarization angle with energy can be tested with respect to a
possible intrinsic polarization angle variability. With an
observation of $10^{6}$~s, values of $\eta$ down to
3~$\times$~$10^{-10}$ can be reached with XIPE using e.g. the
known Blazar 1ES1101$-$232, at z = 0.186, with a clear synchrotron
spectrum and high optical polarization, assuming it has a 10$\%$
polarization degree in the X-ray band. By performing polarization
measurements from several bright enough Blazars at different
distances, observed to pursue other scientific objectives, XIPE
can put the results on a firm statistical basis (as discussed for
GRBs in \cite{Kostelecky2013}).

\subsection{Search for axion-like particles}
Axion-like particles (ALP) are bosons that are predicted in extensions of the Standard Model.
They can form the so-called cold dark matter responsible for the formation of structures in the
Universe and, conversely, be the quintessential dark energy responsible for the acceleration in
the cosmic expansion. Axions mix with photons in the presence of a magnetic field with a rotation of the photon
polarization and possibly with the production of elliptical polarization from linear.

Such an effect was searched on-ground by Polarizzazione del Vuoto
con LASer (PVLAS), a dedicated experiment with initially positive
results but successively withdrawn (\cite{Zavattini2006,
Zavattini2008}). PVLAS has recently been upgraded
(\cite{Zavattini2012}).

On-ground experiments are limited by having short baselines and
the consequently small effects on polarization. Searching for this
effect in distant astrophysical sources overcomes this limitation.
This effect may be detectable in neutron star atmosphere spectra;
however, because photon-axion conversion occurs only for the
O-mode, this signature is much easier to detect using polarization
measurements than standard spectroscopy (\cite{Perna2012}).

\begin{figure}[ht]
\centering
%% Use the relevant command to insert your figure file.
%% For example, with the graphicx package use
\includegraphics [scale=0.5] {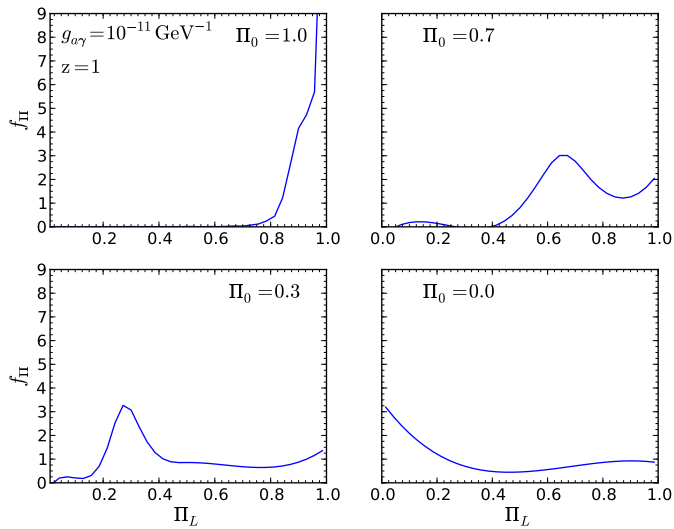}
\caption{Probability density function derived by
simulating the propagation of X-rays in the extragalactic magnetic
field and the photon-ALP coupling. It is shown the final
polarization measured for initially linearly polarized photons
emitted at z = 0.3 from GRBs or other distance sources. The
sources have an initial fixed polarization of 100$\%$, 70$\%$,
30$\%$ or 0$\%$. The magnetic field coherent length scaling and
the plasma density used corresponds to the Wilkinson Microwave
Anisotropy Probe (WMAP) constraints. The figure is from
\cite{Bassan2010}.}
\label{Bassan_fig5}       % Give a unique label
\end{figure}

Other authors (\cite{Bassan2010,Payez2012}) suggested that in case
of a very light ALP, photon-ALP mixing in intergalactic,
intracluster and Galactic magnetic fields may significantly affect
the polarization of radiation emitted by distant sources, inducing
either a linear polarization on initially unpolarized photons or a
dispersion of the degree of polarization of initially linearly
polarized ones (see fig. \ref{Bassan_fig5}).
Clusters of galaxies
emitting in X-rays, and a dozen with a flux large enough for
polarimetry, are expected to have very small or null linear
polarization in origin. A detection of a large polarization from
clusters could be the signature of photon-ALP mixing (see for
example lower-right panel of fig. \ref{Bassan_fig5}). Moreover
imaging allows the contribution of possible AGNs or of foreground
objects in the FoV to be excluded.

Moreover, ALP signatures should strongly depend on energy and on
the projected position of the object on the sky because of the
difference in magnetic field morphology in different directions of
observation. Natural candidates for these studies are again
Blazars where ALP-induced effects can be searched
but also the correlation between the polarization of galactic sources and the
viewing direction. A
sample of Blazars with the synchrotron peak emitting in X-rays, if
the photon-ALP mixing is acting, should show an X-ray polarization
distribution larger with respect to the corresponding distribution
in optical wavelength, due to the presence of a cut-off energy for
this effect, in a way that depends on the distance.

\section{Science requirements for a small imaging X-ray polarimetry mission}
\label{sciencereq} The capability to measure the degree of
polarization and the angle of polarization can be expressed in
terms of the MDP that represents, at a certain confidence level,
the level of the signal which can be attributed solely to
statistical fluctuations in the instrumental response. In fact,
polarization is a positive-definite quantity and therefore it is
always measured to some extent. Only a detection greater than the
MDP is statistically significant, and to reach a 3-$\sigma$
measurement of a particular level of polarization, an integration
time 2.25 longer than that corresponding to the same MDP is
required in case, as
it is common practice in X-ray polarimetry, both the angle and the
modulation amplitude are simultaneously measured
(\cite{Weisskopf2010,Elsner2012,Strohmayer2013}). A measurement at
3-$\sigma$ of the polarization degree allows a 1-$\sigma$
confidence interval on the position angle of about 9.5$^\circ$ to
be reached (\cite{Elsner2012,Strohmayer2013}). If the level of
background is negligible with respect to the counts from the
source, the MDP at 99$\%$ confidence level is expressed as :

\begin{equation}\label{MDP}
    MDP = \frac{4.29}{\mu \sqrt{S} \sqrt{T}}.
\end{equation}

In equation \ref{MDP} $\mu$ is the so-called modulation factor, S
is the source counting rate and T is the observing time. The
requirement on the imaging capability is important for two
reasons: first of all, imaging is necessary to single out the
target source from others in the FoV, thereby reducing the
underlying background. Moreover, imaging is a powerful tool for
performing angularly resolved polarimetry of extended sources
(e.g. Pulsar Wind Nebulae, Supernova Remnants). Since it is
expected that only one solar flare will occur at a time and that
it is usually much brighter than the background, scientific
objectives on solar physics do not pose any imaging requirement.

A moderate energy resolution is required to perform energy-resolved polarimetry of source continua
and to disentangle the dependency on energy of the instrumental response, e.g. the modulation factor
and the efficiency.

XIPE scientific requirements on the timing resolution and timing accuracy are driven mainly by the necessity
to resolve in phase the emission of rapidly spinning millisecond pulsars. The timing requirement for
the solar flares polarimeter is less stringent, except for the dead time.

The pointing accuracy is defined to include in the FoV extended
sources such as the Crab Nebula. The range of duration of an
observation that spans from a few kiloseconds to one week is
requested to arrive at the required sensitivity. However, a set of
on-board calibration sources must be provided to check the
performance stability.

The short duration of the mission implies no particular requirements on the stability performance
since the XIPE payload is already built with space-proven technology typical of X-ray Low Earth Orbit instrumentation.

\section{The payload} XIPE has
four instrument units, two identical Efficient X-ray Photoelectric
Polarimeters (EXPs), a Medium Energy Solar Polarimeter (MESP) and
a Solar Photometer in X-rays (SphinX)
 described in the following paragraphs.

The two EXPs comprise two identical pairs of X-ray telescopes
and focal plane instrumentation. The X-ray
telescopes are a heritage of the JET-X project, for which four
mirror modules (MMs) were developed, as well as three flight
model units (FM) and an Engineering
Qualification Model (EQM) used for the qualification test
campaign but with the same characteristics of the 3 FM units.

%\begin{figure}[ht]
%\begin{center}
%% Use the relevant command to insert your figure file.
%% For example, with the graphicx package use
%\includegraphics [scale=1.3] {GPD_large.eps}
%\caption{The Gas Pixel Detector with a large body.}
%\label{GPD_Large}       % Give a unique label
%\end{center}
%\end{figure}

\begin{figure}[htpb]
\centering
\subfigure[\label{fig:GPD_sketch}]{\includegraphics[scale=3.8]{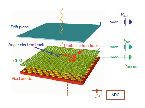}}
\subfigure[\label{fig:GPD_large}]{\includegraphics[scale=0.8]{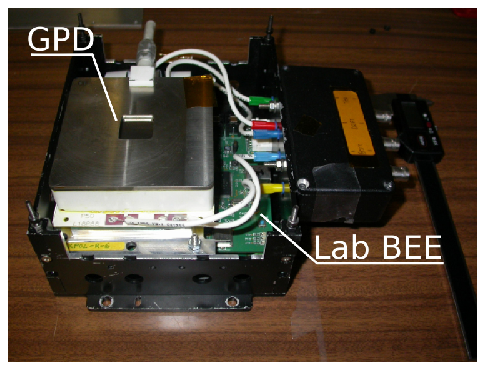}}
%\hspace{1cm}
\subfigure[\label{fig:modfac}]{\includegraphics[scale=5.0]{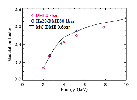}}
\caption{({\bf a}). The sketch of the different components of the
GPD. ({\bf b}) The current GPD prototype having a
larger body, but the same active area, with respect to the first
sealed version (\cite{Bellazzini2007}) with a more uniform
electric field and background. BEE is the laboratory back-end
prototype. The flight BEE is described in paragraph
\ref{electronics&FW}. ({\bf c}).
The measured modulation factor for two filling mixtures, at different energies,
compared to Monte Carlo simulation.} \label{GPD}
\end{figure}

\subsection{The Astronomical X-ray polarimeter}
\label{AstronomPolar} As already mentioned, the XIPE payload is a
descendant of POLARIX (\cite{Costa2010}) but with only two
telescopes, to be compliant with the resources of an ESA small
mission. The Gas Pixel Detector (GPD) is, however, evolved in a
configuration (see fig. \ref{fig:GPD_large}) with a larger cross
section.

\subsubsection{The Detector and the Mirrors}
The GPD (\cite{Bellazzini2007}) is a gas cell made of MACOR filled
with a mixture of 20$\%$ He and 80$\%$ DME, with a thin 50~$\mu$m
Beryllium entrance window glued on a titanium frame, a drift gap
(1~cm), a charge amplification stage and a multi-anode read-out.
X-rays absorbed by the gas are converted into ejected
photoelectrons that in turn produce an ionization pattern in the
gas. The track is drifted by a uniform electric field to the Gas
Electron Multiplier (GEM), and there the charge is multiplied with
a negligible change in the shape.

Below the GEM, at a distance of less than a few hundred $\mu$m,
the top layer of a multilayer ASIC (Application
Specific Integrated Circuit) CMOS (Complementary
Metal$–$Oxide Semiconductor), is covered with 105600 metal
hexagonal pads with 50~$\mu$m pitch and a high filling factor. The
ASIC CMOS (\cite{Bellazzini2006}) chip is glued and internally
bonded to a Kyocera alumina package whose pins are externally
soldered to a Printed Circuit Board. The bottom
layers of the ASIC CMOS constitute complete readout electronics
independent for each pixel. The self-triggering capability is
coupled to the selection capability of the sub-frame containing
the photoelectron track. The image of the latter is analyzed by an
algorithm (\cite{Bellazzini2003, Pacciani2003}) that provides the
impact point with a resolution of 30 $\mu$m rms
(\cite{Soffitta2013}) and an estimate of the emission direction.
Photoelectrons derived from polarized photons have emission
azimuthal directions distributed as cos$^{\mathrm{2}}$($\phi$),
where $\phi$ is the angle with respect to the polarization vector.

While generating the so-called modulation curve, the degree and
the angle of the linear polarization of the incoming photons are
derived from the amplitude and phase.

The mirror modules were developed at the Brera Observatory and
manufactured by Medialario with an electroforming
replica process. They consist of 12 gold coated Nickel shells with
a Wolter I geometry, with diameters going from 191 to 300 mm and a
total length os 600~mm. The focal length is 3500 mm. They have
been calibrated many times at the PANTER X-ray test facility for a
variety of energies and off-axis angles. The total effective area
of a single unit for some energies of interest is reported in
table \ref{tabeffarea}.

The Half Energy Width (HEW) is $\sim$ 15~arcsec at 1.5~keV and
$\sim$ 19~arcsec at 8~keV (\cite{Wells1997}). At the end of
November 2012, as described in sec. \ref{PANTER}, the JET-X (FM 2)
optics was again calibrated at PANTER X-ray test facility.
Preliminary results show (\cite{Spiga2013}) that the effective
area and the angular resolution are presently basically preserved,
and, therefore, the optics are still suitable for an X-ray
experiment with good imaging capabilities. The total mass of each
(MM) unit is of 59.9~kg and the maximum diameter of one mirror
unit is at the interface flange, and this corresponds to a
diameter of 388 mm. A maximum gradient of less than 2 $^\circ$C
assures that the HEW is not degraded by more than 10~arcsec.

JET-X optics have been
already qualified for a launch with Proton rockets (for the JET-X
experiments on-board Spectrum-X-Gamma) and DELTA rockets (for the
Swift mission). The EQM module can be used for the
qualification with other launchers (e.g. Vega). The effective area
of the JET-X optics and the efficiency make the GPD sensitive in
the energy range of 2--10~keV.

The sensitivity achieved with this configuration
(\cite{Muleri2008}) is already compliant with the scientific
requirements. However, a continuous search in the parameter space
of gas thicknesses and mixtures is being conducted with the
intention of arriving at an even better figure of merit (i.e. to
optimize the parameters of the gas mixture, such as the electron
transverse diffusion, the scattering probability and the charge
gain).

A prototype GPD was thermo-vacuum tested between -15$^\circ$C and
+45$^\circ$C and irradiated with a Fe ion dose corresponding to
several years in orbit (\cite{Bellazzini2010}). GEMs produced with
the current technology have been successfully irradiated with
protons and heavy ions (\cite{Iwashashi2011}). The main
characteristics of the EXPs GPD on-board XIPE are summarized in
table \ref{tabexp}.

\begin{table}[ht]
\caption{Characteristics of EXP and MESP on-board XIPE}
\begin{tabular}{ll}
\hline\noalign{\smallskip}
Parameter & 2 $\times$ EXP units (2 $\times$ MESP units)  \\
\noalign{\smallskip}\hline\noalign{\smallskip}
Polarization sensitivity & MDP = 14$\%$ in 100 ks for 1 mCrab \\
                         & 0.6$\%$ for an X10 flare; 6.6$\%$ for an X5.2 flare (MESP, 15-35~keV)\\
Imaging capability & 24 arcsec (HEW, overall), 14.7 $\times$ 14.7 $arcmin^{2}$ FoV \\
Spectral resolution & 20$\%$ @ 5.9~keV \\
Timing & Resolution: 8 $\mu$s; Accuracy: 2~$\mu$s;  Dead time: 10 $\mu$s (negligible)  \\
Gas mixture & 20$\%$ He-80$\%$ DME 1-atm 1-cm (EXP)\\
        & 60$\%$ Ar-40$\%$ DME 3-atm 3-cm (MESP)\\
Energy range & 2--10 keV (EXP) \\
             & 15--35 keV (MESP) \\
Background & EXP: 5.5 $\times$ $10^{-7}$ c/s (4.8 nCrab, point source) (\cite{Soffitta2012})\\
           & MESP: negligible for solar flare X-ray polarimetry \\
\noalign{\smallskip}\hline
\end{tabular}
\label{tabexp}       % Give a unique label
\end{table}

It should be noted that the scientific requirements for XIPE are fulfilled even with a payload
hosting a single GPD and a single telescope. Two telescopes allow the same
sensitivity to be obtained in half of the time, making possible larger population studies in the time
frame of the mission.

\subsubsection{The background of XIPE}
\label{background}
The residual background of the GPD is not, at the present stage of development, minimized
by means of the use, for example, of an anti-coincidence system or pulse shape discrimination
as in the case of a traditional gas multiwire proportional counter. However, all these methods
exploit in some way the space distribution of the charges produced by the detection event. The same and possibly
better can be done with the GPD that resolves the track.
The high granularity of the detector surface and simple
methods of pattern recognition allow a very low and uniform background to be reached.
Very briefly, we show here what kind of background rejection can be applied :

\begin{itemize}

    \item \textbf{Amplitude}. The spectroscopic capability of the GPD allows for setting a lower and an upper
          energy threshold.
    \item \textbf{Maximum window}. The ASIC CMOS chip can be configured setting a maximum allowed window frame.
          Background events are characterized by a window frame larger than that of an X-ray event.
    \item \textbf{Number of pixels}. X-rays provide a smaller non-zero pixel number with respect to the minimum
          ionizing background electrons.
    \item \textbf{Contiguity of the track}. Minimum ionizing particles produce tracks that can be discontinuous
          while this case is much more infrequent for higher ionizing particles such as X-ray photoelectrons.
    \item \textbf{Difference in skewness}. Background tracks due to minimum ionizing particle are characterized by a more
          uniform charge distribution with respect to photoelectron tracks, with a consequent difference in skewness.
\end{itemize}

Due to the finite range of the electrons, the most external pixel
frame of the ASIC CMOS chip cannot be easily used because not all
the azimuthal angles are detected with the same coverage. Such a
frame represents a sort of side-anticoincidence that allows for
excluding the background events arriving from the four sides. With
the current GPD (see fig. \ref{fig:GPD_large}) having a larger
body with respect to that proposed for POLARIX, the background
does not accumulate close to the ASIC edges.

Applying the above prescriptions, we expect that the background
rate in orbit is conservatively well-estimated by means of past
experiments with gas detectors filled with similar mixtures. We
therefore evaluated the background of the GPD by using the
measurements reported in \cite{Bunner1978} for the gas detector
filled with Ne-$CO_{2}$ and extrapolating the results from the
$CH_{4}$ gas detector but taking into account the difference in
number of electrons in the molecules. The background estimates are
shown in table \ref{tabback}.

\begin{table}
\caption{Expected residual background for XIPE EXP.}
\begin{tabular}{lp{1.8cm}p{1.3cm}p{1.5cm}p{2.1cm}p{1.8cm}}
\hline\noalign{\smallskip}
Source & Extension & Source rate & Diff. Backg. & Resid. Ne-$CO_{2}$ (\cite{Bunner1978})  & Resid. $CH_{4}$ (\cite{Bunner1978}) \\
       &           &  (c/s)  & (c/s) & (c/s)                               &  (c/s)                          \\
\hline\noalign{\smallskip}\hline\noalign{\smallskip}

Point-like & 24'' HEW            & 2 $10^{-3}$-200 & 5 $10^{-12}$ & 5.5 $10^{-7}$ & 5.6 $10^{-6}$ \\
           &     (407 $\mu$m)    &                 &              &  (4.8 nCrab)  & (49 nCrab) \\
\hline
SgrB2 & 1.5' $\times$ 3.5'    & 5.8 10$^{-4}$     & 1.5 $10^{-10}$ & 1.8 $10^{-5}$ & 1.9 $10^{-4}$ \\
      & 1.5 $\times$ 3.5 $mm^{2}$ & (5 $\mu$ Crab)   &            & (160 nCrab)   & (1.6 $\mu$Crab)\\
\hline
\end{tabular}
\label{tabback}       % Give a unique label
\end{table}

\subsection{The solar X-ray polarimeter and photometer}
\label{SolarPolarimeter} Polarimetry of solar flares is still a
debated unresolved issue in astrophysics notwithstanding
experiments launched since the beginning of X-ray astronomy. We
decided to include in the payload two Medium Energy Solar Flare
polarimeters (MESPs) designed with the same technology as the low
energy polarimeters but filled with an Ar-DME (60-40) mixture. The
drift region is 3-cm thick. The two MESPs always
face the sun within an accepted angle of $\pm$ 30$^\circ$
depending on the pointing constraints of the EXPs. Their open
configuration is equipped with a field-of-view angular delimiter
(FAD) that reduces the X-ray background, while the insertion of
a multi-layer gray filter suppresses the low-energy,
low-polarization, large photon flux expected during flares.
The acquired data are continuously stored in a
cyclic memory and, after a positive on-board trigger condition is
verified, they are saved to be downloaded on-ground. The two
MESPs are effective in the 15--35~keV energy range. At these
energies, the intensity of the flare is still large, and a small
geometrical area provides sufficient sensitivity. Also, the
non-thermal bremsstrahlung starts to dominate the emission with an
expected high degree of linear polarization.

Notwithstanding the intrinsic good imaging capability of MESP, no
information on the location of the flare with respect to the solar
limb can be directly derived by this open-sky configuration. The
modelling of the expected polarization degree will, therefore,
need information from other solar missions. The MESP design has
already been developed for the New Hard X-ray Mission (NHXM,
\cite{Tagliaferri2012}) project and is validated in laboratory
with a prototype 2-cm 2-atm thick filled with a mixture Ar-DME
70-30 (\cite{Fabiani2012}). To complement the study of the flare,
an X-ray photometer, SphinX, uses a silicon PIN detector for high
time resolution (10~$\mu$s) measurements of the solar spectra of
quiet and active corona in the range 1.2--15~keV. The SphinX
instrument is a heritage of CORONAS-Photon payload and its volume
is 27~$\times$~7~$\times$~22~cm$^{3}$. A new more compact and
lighter design is under study to be included in future missions
(\cite{Sylwester2008}).

\paragraph{Inclined penetration effects for the solar polarimeter}
   The requirement on the
  sky visibility for EXP means that the pointing direction of MESP with respect
  to the Sun is $\pm30^\circ$. Photons coming from flares could therefore impinge
  on the detector at an inclined angle. This effect has been studied for a GPD with a He-DME
  mixture in \cite{Muleri2010b} and is the object of a forthcoming paper (\cite{Muleri2013}).
  In \cite{Muleri2010b}, it has been shown that when applying the standard analysis to the modulation curve,
  prominent systematic effects are observed which can, however, be
  precisely corrected with the procedure described in the same paper. Moreover,
  recent laboratory measurements by The Space Research Center in Poland also showed that, with this inclination,
  SphinX is still sensitive for observing the
  emission from the Sun.

\subsection{The electronics and the filter wheel}
\label{electronics&FW} In the vicinity of each of the two GPDs of
the EXP and of both of the two GPD of the MESP the three Back-End
Electronics (BEE) are located. Each BEE is responsible for
distributing the low voltages to the ASIC CMOS, controlling, by
means of a dedicated Field-Programmable Gate Array
(FPGA), the ASIC CMOS, performingAnalog-to-Digital
(A/D) conversion of the ASIC output signals, as well as the
zero-suppression, time-tagging each event with an accuracy of
2~$\mu$s, and, finally, implementing a Peltier Driver 'for the GPD
temperature control to be compliant with the GPD stability
requirement ($\pm$ 2$^\circ$C within +5$-$+20 $^\circ$C). Each GPD
requires three high voltage power supply lines (HV) in the range
0.2$-$3~kV (EXP) and 0.2$-$6~kV (MESP) and currents of a few
nanoamperes.

Two filter wheels (FW \#1 and FW \#2), whose design is derived
from that successfully flying on board {\it XMM-Newton}, are
placed in front of each of the two EXP GPDs. Each observation mode
corresponds to a different FW position. Each position allows for
optimizing polarimetry in case of bright or multiple sources in
the FoV, for calibrating the gain and the response to polarized
X-rays (\cite{Muleri2007}) and for the gathering of the internal
background. The GPD, the BEE and the FW compose the separated
Focal Plane Assembly (FPA) of each EXP MM.

\subsection{The Interface Electronics} XIPE takes advantage of the
\label{InterfaceElectronics} spacecraft On Board Data Handling
unit (OBDH) in order to share with the bus some of the most
relevant functionality such as the data-storing and the
preparation of packets, at variance with POLARIX with its
dedicated Payload Data Handling Unit (PDHU).

At the payload level, there is  an Interface Electronics (I/FE)
with the remaining functionalities. The I/FE is responsible for
configuring the BEEs of EXP, MESP and SphinX,
including the provision of regulated Low Voltages
(LV) from the unregulated power provided by the solar panels.
 generating and managing the housekeeping, being in charge of the
EXP FWs, taking care of the non regulated primary power bus
providing the secondary voltages needed by the units, managing the
Pulse Per Second synchronization signal line, parsing and
executing the telecommands coming from the spacecraft, and
managing the Payload Instrument Operative Modes (Boot,
Maintenance, Idle, Observation and Test).

\begin{figure}[ht]
\begin{center}
%% Use the relevant command to insert your figure file.
%% For example, with the graphicx package use
\includegraphics [scale=0.5] {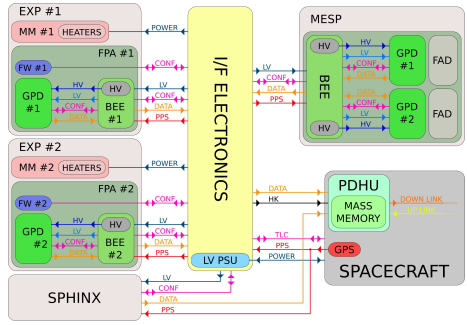}
\caption{The Block Diagram of the XIPE experiment.
Here FW is the Filter Wheel (one for each of the two
EXPs) that includes filters, unpolarized and polarized calibration
sources; LV is the regulated low voltage that powers the three
Back End Electronics (BEE) that include also the high voltages
(HV), and manage the GPDs; MM is the X-ray Mirror Module (two MMs
in total), FAD (field angular delimiter) is the mechanical shield
for the Cosmic X-ray Background to be used in front of the solar
X-ray flares polarimeter; PDHU is the payload data handling unit
hosted in the spacecraft and incorporating the XIPE mass
memory.}
\label{blockdiagram}       % Give a unique label
\end{center}
\end{figure}

\section{Some results from the GPD-JETX mirror calibration at the PANTER X-ray test facility}
\label{PANTER}

\begin{figure}[ht]
% Give a unique label
\begin{center}
%% Use the relevant command to insert your figure file.
%% For example, with the graphicx package use
\includegraphics [scale=0.3] {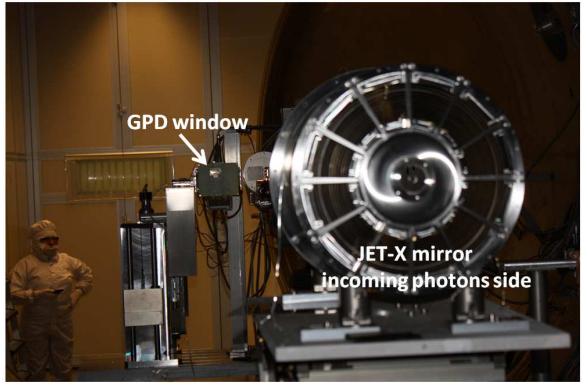}
\caption{The JET-X mirror at the front and the GPD in the
background in the experimental chamber at the PANTER X-ray test
facility.} \label{fig:PANTER_FOTO}
\end{center}
\end{figure}

The block diagram of the XIPE payload is shown in fig.
\ref{blockdiagram}, here we show the first calibration of an X-ray
polarimeter with X-ray optics, namely one of the two flight models
of JET-X, has been performed at the Max Planck Institute for
Extraterrestrial Physics PANTER X-ray test facility at Neuried by
Munchen (Germany), see fig. \ref{fig:PANTER_FOTO}. A stand-alone
new calibration of the JET-X optics was performed with very good
results, implying that JET-X optics are still suitable for an
X-ray space experiment. Here, we present the preliminary
re-measurement of the effective area (see table. \ref{tabeffarea})
compared with theoretical expectations.

\begin{figure}[ht]
\begin{center}
%% Use the relevant command to insert your figure file.
%% For example, with the graphicx package use
\includegraphics [scale=0.5] {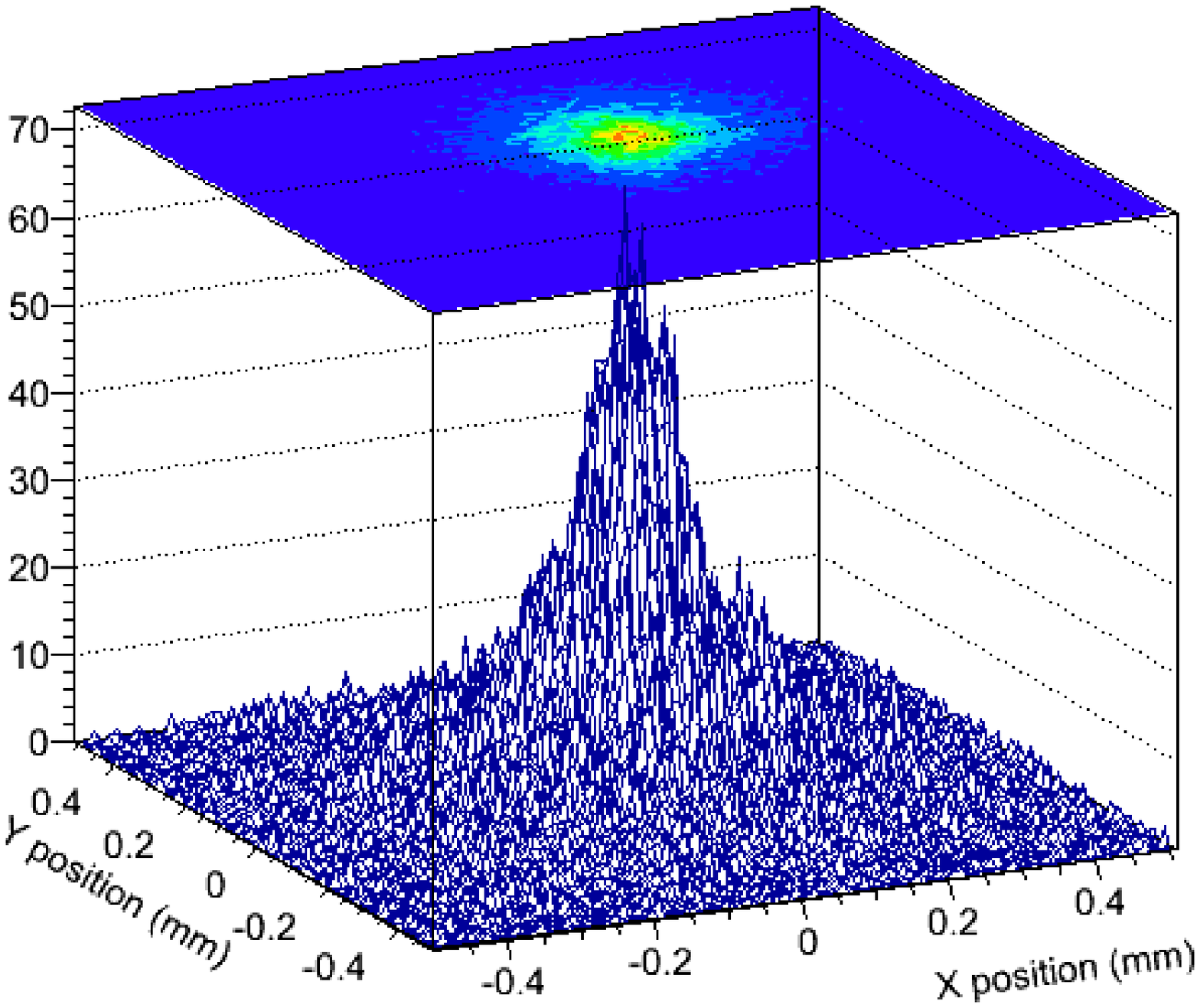}
\caption{Image of the PANTER X-ray source at 4.5~keV, focused by
the JET-X optics and imaged by the GPD. The Z axis represents the
total counts per bin. The focal length of the
JET-X optics for a point source at infinite is 3500 mm while the
focal length for the finite distance at PANTER X-ray facility is
3600 mm. The Half Energy Width measured at PANTER in this
configuration corresponds to 23.2~arcsec while the plate scale is
about 1 arcmin/mm.} \label{JETX_GPD_image}
\end{center}       % Give a unique label
\end{figure}

By means of Monte Carlo simulations and measurements, we already
showed that the overall angular resolution is dominated by the
point spread function of the optics and, secondarily, by the
inclined penetration in the gas drift thickness
(\cite{Lazzarotto2010, Soffitta2013}). In this regard, we measured
(\cite{Fabiani2013}) at the PANTER X-ray test facility a position
resolution, at 4.5~keV, of 23.2~arcsec, fully consistent with the
estimate done at this energy. In fig. \ref{JETX_GPD_image} we show
the image of the X-ray source at 4.5~keV focused by the optics and
detected by the GPD.

\begin{table}[ht]
\caption{On-axis theoretical and measured effective area during
the November 2012 calibration campaign at the PANTER X-ray test
facility.}
\begin{tabular}{lll}
\hline\noalign{\smallskip}
Energy  & Measured effective area  & Theoretical effective area \\
  (keV) &  (cm$^{\mathrm{2}}$)  & (cm$^{\mathrm{2}}$) \\
\noalign{\smallskip}\hline\noalign{\smallskip}
2.99 & 109 & 105\\
4.54 & 112 & 110\\
6.4 & 96 & 104\\
8.04 & 53 & 61\\
\noalign{\smallskip}\hline
\end{tabular}
\label{tabeffarea}       % Give a unique label
\end{table}

We also performed a dedicated long run to arrive at a useful upper
limit on the spurious modulation introduced by the optics at
4.5~keV. The results of this analysis, that require a careful
evaluation of the modulation from the underlying bremsstrahlung,
will be presented in a forthcoming paper. We plan in the future to
perform an end-to-end calibration with the use of a polarized
source at the same facility.

\section {The Payload budgets}
The XIPE telemetry budget is reported in table \ref{telemetrytab}. The
typical data rate for the two EXP units observing a typical source (with
a flux of 200 mCrab) is 29 kbit/s, plus the housekeeping (4 kbit/s
in total). For the purpose of sustaining high data rates in the case
of bright sources, two strategies can be envisaged in the proposed
architecture. The baseline is the temporary data storage in the
PDHU mass memory, which is part of the spacecraft, combined with alternating between
bright and faint sources in the observation sequence so that the average data
rate is below that sustainable by the downlink. In this scenario,
the PDHU mass memory dedicated to the two EXP units must be $>$2~GB to store the data collected in a observation lasting $10^{5}$~s
of a bright source such as the Crab Nebula. Alternatively, the PDHU
can perform in real-time the track reconstruction, and transmit on
ground only the main results of the analysis. MESP data will be
continuously transferred and stored in the PDHU but they will be
transmitted to Earth only if a flare is detected. This is
expected to occur sporadically: about 20 solar flare events
lasting a few tens of minutes will be observed by the MESP during
the mission lifetime of 2 years. In case of a trigger,
the data starting from 5 minutes before the trigger will be
completely downloaded, that is, without real-time on-board
analysis, together with EXP data. The PDHU mass memory dedicated
to MESP must be at least 2.5~GB to completely store the data collected
for a X10 class flare (the brightest kind of event) lasting 30
minutes. The telemetry requirement for the SphinX instrument is 8~kbit/s.

The XIPE payload mass budget is reported in table \ref{tabMass}.
XIPE is directly derived from POLARIX phase A and also benefits
from a study for IXO, a proposed ESA large mission. The payload
mass budget is therefore well-known. The mass of the flight
telescopes, which contribute a large part of the mass, is known
because they already exist. The GPD prototypes currently in use
are already built with flight components; therefore, their (low)
mass can be evaluated with high accuracy. The associated
electronics for the XIPE detector payload has a standard design.
Therefore, a definite evaluation of the mas of the flight hardware
is possible. The payload structure is the large mounting plate,
indeed part of the payload, where the mirrors, the detectors and
their electronics are fixed.

Finally, the XIPE power budget is shown in table \ref{tabPower}. A
large fraction of the power is for the thermal
balancing of the Mirror Modules in order to prevent possible distortion due to
temperature gradients. Taking into account that one JET-X
telescope is currently flying on the Swift satellite, that the
lower energy threshold is larger (2~keV for XIPE) and finally that
the overall angular resolution is also determined
by the inclined penetration in the GPD, the requested
power for the telescope thermal stability is known, and is less demanding than that
of XRT-SWIFT.

\begin{table}
\caption {XIPE Telemetry budget}
\begin{tabular}{ll}
\hline
  2 $\times$ EXP units & 29~kbit/s (typical ) \\
  \hline
  MESP & $<$ 2.5 GB (sporadically, once per month) \\
  SphinX & 8~kbit/s  \\
  HK & $<$ 4~kbit/s \\
  \hline
\end{tabular}
\label{telemetrytab}
\end{table}

\begin{table}
\caption{The Payload Mass budget of XIPE. CBE is
the Current Best Estimate and DMM is the Design Maturity Margin
(it includes an additional cautionary percentage depending on the
maturity level of each items, FAD is the Field Angular Delimiter
for the solar polarimeter (MESP).}
\begin{tabular}{lllllll}
\hline\noalign{\smallskip}
 & & & PAYLOAD MASS BUDGET  & & & \\
 & No. of Item & CBE &  & DMM & CBE+DMM \\
 & & Mass(kg) & Tot (kg) & & Mass (kg) & Mass (kg)\\
\noalign{\smallskip}\hline\noalign{\smallskip}
\textbf{EXP} & & & & & & \textbf{203.6}\\
MM & 2 & {\color{blue} 60\dag} & 120.0 & 5\% & 126.0 & \\
MM baffles + heater + harness & 2 & {\color{blue} 5.0\dag} & 10.0 & 5\% & 10.5 & \\
MM mounting structure & 1 & 10.0 & 10.0 & 20\% & 12.0 & \\
GPD+FW+Baffle+mech. \emph{\textbf{interf.}} & 2 & {\color{red} 3.3\ddag} & 6.6 & 10$\%$ & 7.3 &\\
BEE+mech \textbf{\emph{interf.}} & 2 & {\color{red} 1.6\ddag} & 3.1 & 10\% & 3.4 &\\
Payload structure & 1 & 32.0 & 32.0 & 20\% & 38.4 & \\
Sun Shield & 1 & 5.0 & 5.0 & 20\% & 6.0 & \\
\textbf{MESP} & & & & & & \textbf{16.0} \\
GPD+mech \textbf{\emph{interf.}} & 2 & 3.0 & 6.0 & 10$\%$ & 6.6 & \\
BEE+mech \textbf{\emph{interf.}} & 1 & 2.0 & 2.0 & 10$\%$ & 2.2 & \\
FAD & 2 & 3.0 & 6.0 & 20$\%$ & 7.2 & \\
\textbf{SPHINX} & & & & & & \textbf{3.7} \\
All integrated instrument & 1 & 3.5 & 3.5 & 5$\%$ & 3.7 & \\
\textbf{Other} & & & & & & \textbf{10.7}\\
StarTracker & 1 & 1.0 & 1.0 & 5$\%$ & 1.1 & \\
I/FE & 1 & 5.0 & 5.0 & 20$\%$ & 6.0 & \\
Harness & 1 & 3.0 & 3.0 & 20$\%$ & 3.6 & \\
 & & & & & & \textbf{TOTAL} \textbf{234.0} \\
{\color{blue}\dag POLARIX heritage} & & & & & &\\
{\color{red}\ddag IXO heritage} & & & & & &\\
\noalign{\smallskip}\hline
\end{tabular}
\label{tabMass}       % Give a unique label
\end{table}

\begin{table}
\caption{The Payload Power budget of XIPE}
\begin{tabular}{lllllll}
\hline\noalign{\smallskip}
 & & & PAYLOAD POWER BUDGET  & & &\\
 & No. of Item & CBE &  & DMM & CBE+DMM \\
 & & Power(W) & Power (W) & & Power (W) & Power (W)\\
\noalign{\smallskip}\hline\noalign{\smallskip}
\textbf{EXP} & & & & & & \textbf{168.0}\\
MM thermal control (peak)& 2 & \textcolor[rgb]{0.00,0.50,0.25}{50\dag} & 100.0 & 20\% & 120.0 & \\
GPD & 2 & {\color{red} 2.0\ddag} & 4.0 & 20\% & 4.8 & \\
BEE & 2 & {\color{red} 12.0} & 24.0 & 20\% & 28.8 & \\
FW (peak) & 2 & {\color{red} 6.0} & 12.0 & 20$\%$ & 14.4 &\\
\textbf{MESP} & & & & & & \textbf{28.8}\\
GPD & 2 & 2.0 & 4.0 & 20\% & 4.8 &\\
BEE & 1 & 20.0 & 20.0 & 20\% & 24.0 & \\
\textbf{SPHINX} & & & & & & \textbf{15.8} \\
All integrated instrument \\ (peak) & 1 & 15 & 15.0 & 5$\%$ & 15.8 & \\
\textbf{Other} & & & & & & \textbf{29.5}\\
StarTracker & 1 & 6.0 & 6.0 & 5$\%$ & 6.3 & \\
I/FE & 1 & 19.3 & 19.3 & 20$\%$ & 23.2 & \\
     &   &      &      &        &      & \textbf{TOTAL 242.1} \\
{\textcolor[rgb]{0.00,0.50,0.25}{\dag Swift heritage}} & & & & & &\\
{\color{red}\ddag IXO heritage} & & & & & &\\
\noalign{\smallskip}\hline
\end{tabular}
\label{tabPower}       % Give a unique label
\end{table}

\begin{table}
\caption{System constraints of XIPE}
\begin{tabular}{ll}
\hline\noalign{\smallskip}
Constrain & Value   \\
\noalign{\smallskip}\hline\noalign{\smallskip}
Satellite stabilization & Three axis \\
Absolute Pointing Error & 3 arcmin \\
Absolute Measurement Accuracy & 10 arcsec, 5 Hz \\
Sky accessibility & $90^\circ$ $\pm$ $30^\circ$ \\
Average scientific telemetry rate & 50 kbit/s \\
Minimum size storage & 5 GByte \\
\noalign{\smallskip}\hline
\end{tabular}
\label{tabconst}       % Give a unique label
\end{table}

\section{Mission Profile}

\subsection{Proposed launch vehicle}
The mission concept is based on a launch into a circular, low equatorial orbit by the Vega Launcher.
The orbit considered for XIPE is of Low Earth Orbit (LEO) type, and has  characteristics reported
in table \ref{tablaunch}.
\begin{table}[ht]
\caption{Requirements on the launcher.}
\begin{tabular}{ll}
\hline\noalign{\smallskip}
Parameter & Requirement \\
\noalign{\smallskip}\hline\noalign{\smallskip}
            Altitude & 600 $\pm$ 16 km for two years mission and controlled re-entry. \\
            Inclination  & $5^\circ$ $\pm$ $1^\circ$ \\
            Eclipse duration & 36 minutes max \\
            Ground Station & 8-11 min contact \\
\noalign{\smallskip}\hline
\end{tabular}
\label{tablaunch}       % Give a unique label
\end{table}

\subsection{Mission duration}
The nominal mission duration is 2 years, plus 1 month for commissioning, 3 months for the Science
Verification Phase (SVP) and 1 month for decommissioning. Such a duration is sufficient to address the science goal
presented in section \ref{sec:Astroph}.

\subsection{Ground Station}
The satellite in nominal operation phase is supported by a
dedicated low latitude ground station, such as Malindi, during its
entire lifetime. The Malindi ground station is optimally located
for the near equatorial XIPE-satellite orbit. The coverage pattern
for this LEO altitude is a regular sequence of contacts, 15 per
day, once per orbit, each one followed by a gap of about 85 min.
Assuming this contact time and a telemetry data rate as high as
512~kbps (included Reed-Solomon on-board coding {for error
correction managing), the downloadable data volume is 3.25~Gb/day.

\subsection{Communication requirements}
The communication requires, as a minimum, the use of S-Band
communication system, which is available at Malindi Ground
Station. This is made compatible with the mission goals by means
of alternating the observation of bright source with subsequent
observation of dim sources, coupled with storage of the data in
the on-board memory.

\subsection{Ground Segment}
The XIPE Ground Segment (G/S) performs the main functions/operations needed at ground level to manage
the mission in terms of both satellite control and global data management. The planned G/S includes the
ground station of the Italian Space Agency (ASI) located at Malindi and the Mission Operation Center (MOC)
of INPE at S\~{a}o Jos\'{e} dos Campos (Brazil). The G/S  monitors and controls
the satellite platform and payload, performs the orbit/attitude operations and generates the orbital
products used for satellite Monitoring $\&$ Control, for the payload management and mission planning.
The G/S generates the mission planning and checks, according to scientific observation requests coming from the User Segment
(U/S). It acquires the raw satellite data (housekeeping and telemetry) and it transfers them to the U/S for processing.
The G/S also includes the Satellite Simulator and the Communication Network responsible for interconnecting the Ground Segment
facilities and providing the related communication services in a secure and reliable way.

The U/S manages the scientific observation requests coming from the scientific community,
forwards them to the G/S for payload scheduling activities, ingests the raw satellite data coming from the G/S, and
generates, archives, catalogues and delivers the scientific data and the data products to the user community.

\subsection{Alternative mission scenario}
The utilization of Iridium Next, at variance with the bus proposed
for POLARIX, also gives the opportunity for significative launch
cost reduction by means of participation in twin or a dedicated
launch with DNEPR also called SS18), an ukraine launch vehicle
named after the Dnieper River, with a more favorable cost with
respect to Vega. This alternative has not been studied so far. It
would imply a highly inclined orbit. The expected background would
be larger in this case, but we should not forget that its counting
rate is smaller than the source counting rate by orders of
magnitude for any observation with realistic observing time for
X-ray polarimetry.

\section{System requirements and spacecraft key factors}
The key constraints on the satellite characteristics are shown in
table \ref{tabconst}. The photoelectric X-ray polarimeters are
based on GPD technology, a non-dispersive instrument with
intrinsic homogeneous azimuthal response irrespective of the
conversion point of the photons in the gas. Therefore, rotation is
not required since the systematics induced by grazing incidence
X-ray optics are not expected to provide additional instrumental
contribution unless down to a very low level (\cite{Almeida1993}).
This fact facilitates the use of platforms with the usual
three-axis stabilization. The requirement on the Spacecraft
Absolute Pointing Error is driven by the FoV of the instrument and
by the requirement on the misalignment of the two optics. The
requirement on the Absolute Measurement Accuracy allows for not
degrading the PSF of the instruments and it exploits the
capability of the GPD to measure the conversion point with high
precision and the photon-by-photon transmission on-ground. The
XIPE science data down-link and storage capability are compatible
with S-band to observe an average source.  A storage of 5~GByte
allows the acquisition of 24 hours of data from Crab plus an X-ray
flare of class X10.

Some operations, usually performed at payload level, are
in this case demanded of the hardware installed at bus (platform) level.
This is possible because of the characteristics of the proposed
platform.
The OBDH acquires and stores the science
data, taking care of the satellite telecommands, of the telemetry
and of the attitude control (AOCS). The thermal control, including
that of the telescopes and of the detectors, is also performed
by the OBDH.

The on-board processor, with associated electronics and memory,
decodes and distributes the payload telecommand. It processes, if
needed, the data evaluating the impact point and the photoelectron
emission angle, and packets the data and the Housekeeping, storing
them in the OBDH memory. Finally, it manages the Pulse Per Second
synchronization signal (for the On-Board Timing(OBT) and Universal
Time (UT) synchronization) line by using the GPS system. The
accuracy and precision of the on-board clock should be such to
time tag the event with 2~$\mu$s accuracy.

\section{The platform for XIPE}
The  size and pointing requirements for the XIPE telescopes are
the drivers for the selection of the XIPE service module. In this
framework, the utilization of the Iridium Next platform, developed
by Thales Alenia Space firm on behalf of Iridium Communication
Inc., is very promising. This product  is now commercialized
with the name of \emph{Elite} platform. The
advantages of this three-axis stabilized platform is in the cost
resulting from the mass production of 81 satellites. Also, the
mass production fits perfectly the launch date in 2017. The
utilization of this kind of platform for an X-ray mission is not a
novelty because it has been already considered for other science
missions.

The Iridium Next platform is equipped with a 2.0~kW solar array
mounted on a double articulated arm. The high modularity of the
platform allows for the ``plug $\&$ play'' of the payload module.
By means of a preliminary payload accommodation study, a XIPE
satellite concept based on \emph{Elite} platform was developed in
collaboration with Thales Alenia Space Italia. This concept is
shown in figure \ref{fig:XIPE_payload}.

Iridium Next can also provide an option based on X-band down-link
transmission and a propulsion capability, allowing the orbit maintenance
(if necessary) as well as the management of satellite end-of-life.
The Iridium Next platform and the Vega launcher can be matched by means
of a special adapter. The accommodation of XIPE within the fairing
of the VEGA launcher is shown in fig. \ref{fig:XIPE_fairing}.

\begin{figure}[ht]
\begin{center}
%% Use the relevant command to insert your figure file.
%% For example, with the graphicx package use
\includegraphics [scale=0.3] {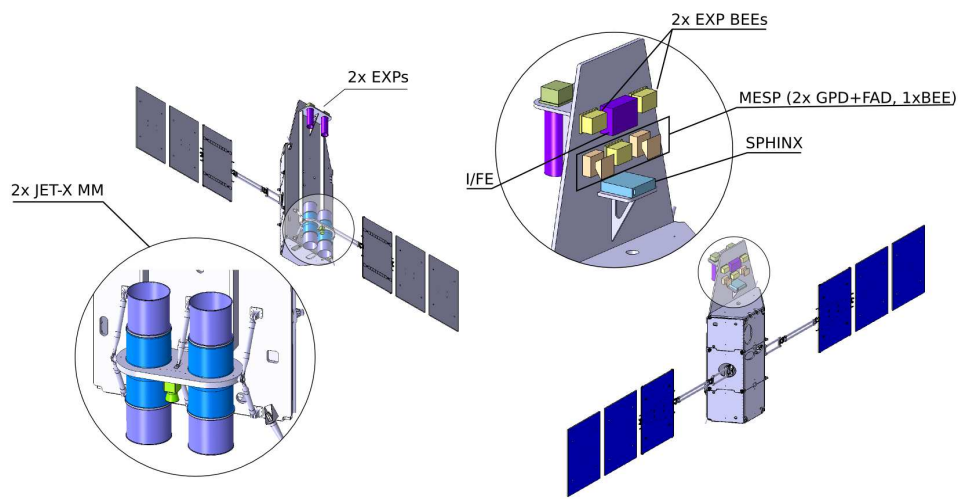}
\caption{XIPE payload accommodated in the Iridium NEXT bus.}
\label{fig:XIPE_payload}       % Give a unique label
\end{center}
\end{figure}

\begin{figure}[ht]
\begin{center}
%% Use the relevant command to insert your figure file.
%% For example, with the graphicx package use
\includegraphics [scale=0.2] {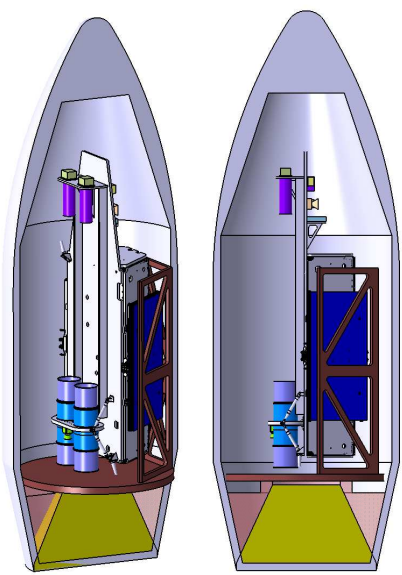}
\caption{XIPE satellite in the fairing of the VEGA bus.}
\label{fig:XIPE_fairing}       % Give a unique label
\end{center}
\end{figure}

The platform and the payload can also be concurrently
integrated and tested at a very late stage, reducing interference
and therefore optimizing the schedule. The Iridium platform mass
is about 450 kg and the power consumption is about 500 W.

\section{Science operation and archiving}
The XIPE science operation and archiving are different from those of standard ESA missions. The team owns the
Science Operation Phase data, lasting three
months, and the Core program data for 25$\%$ of the net observing time. The remaining 75$\%$ net observing
time is assigned to a peer-reviewed Guest Observer program. Target of Opportunity observations are possible and a
proprietary period of one year is guaranteed.

The data are, after a proper check and standard processing,
delivered to the owner. They are in the form of a photon list
containing the time, the absorption point, the energy and the
emission angle, plus the data on coverage, time windows and dead
time. Specific nonstandard analysis is possible by means of a
software package distributed and documented by the
team of the Scientific Data Center of the Italian Space Agency
(ASDC), that also stores all the data and products at their
different steps. Data format and the calibration
database (CALDB) are written in OGIP (Office of
Guest Investigator Programs)$-$FITS (Flexible Image Transport
System), that is the standard file format in X-ray astronomy,
with a total expected amount of 5 terabyte. The Science Operation
Center is located at the INAF/IAPS, INAF/OAB, INFN/Pisa and
ASDC$-$Frascati.

\section{Conclusions}
The most recently flown dedicated X-ray polarimeter dates back to
the 1970s. Motivated by the interest among theoreticians for
opening a new window in X-ray astronomy, where almost all the
classes of sources are expected to be polarized, many missions
with on-board X-ray polarimetry have been proposed since then, and
some arrived at various levels of completion. These include the
fully tested and calibrated Stellar X-ray
Polarimeter (\cite{Kaaret1989, Kaaret1994, Tomsick1997,
Soffitta1998}) aboard the former and not flown Spectrum X-Gamma
Russian satellite, and GEMS, that was discontinued by NASA in May
2012.

In this paper, we described XIPE, proposed in June 2012 for an ESA
small mission aimed at performing spectral-imaging polarimetry of
celestial sources and of solar flares. XIPE takes advantage of
already existing X-ray optics with good imaging capabilities and
an already existing improved version of the GPD X-ray polarimeter,
built with materials and techniques already suitable for use in
space. XIPE is a low-risk mission with a limited need for
resources. The MDP that XIPE can reach (14$\%$ at 1 mCrab in
10$^{5}$ s of observing time in 2$-$10~keV energy band) is
compliant with the polarization expected by most of the classes of
galactic sources and with a limited sample of luminous AGNs.

The imaging capability allows for resolving polarimetry of the
Crab and of other PWN and supernova remnants, disentangling the
presence of multiple sources in the FoV and maintaining the
background at a level compatible with making all observations
source dominated. The feasibility of XIPE has been further
demonstrated by having tested the proposed JET-X optics, either
stand-alone or with the GPD gas detector in its focus, in a
dedicated calibration campaign at the PANTER X-ray test facility.

\section{Acknowledgments}
This work is partially funded by ASI, INAF and INFN.

\bibliographystyle{spbasic}      % basic style, author-year citations
\bibliographystyle{spmpsci}      % mathematics and physical sciences
\bibliography{references}   % name your BibTeX data base

%\begin{thebibliography}{}
% etc
%\end{thebibliography}

\end{document}